\documentclass[twocolumn]{aastex631}

\usepackage{enumerate}
\usepackage{graphicx}
\usepackage{amsmath}
\usepackage{float}
\usepackage{xcolor}
\usepackage{hyperref}
\usepackage{mhchem}
\hypersetup{
    bookmarks=true,                 
    unicode=false,                  
    pdftoolbar=true,                
    pdfmenubar=true,                
    pdffitwindow=true,              
    pdfstartview={FitH},            
    pdftitle={Nuclear uncertainties}, 
    pdfauthor={Mumpower },          
    pdfsubject={Astrophysics},      
    pdfcreator={dvipdf},            
    pdfproducer={dvipdf},           
    pdfkeywords={r-process, stars}, 
    pdfnewwindow=true,              
    colorlinks=true,                
    linkcolor=magenta,              
    citecolor=teal,                 
    filecolor=magenta,              
    urlcolor=cyan,                  
    breaklinks=true,
    linktocpage
}

\graphicspath{{./}{figs}}
\newcommand\Msun{M$_\sun$}

\revised{\today}
\submitjournal{\apj}

\shorttitle{Let there be neutrons!}
\shortauthors{Mumpower et al.}

\begin{document}

\title{Let there be neutrons! Hadronic photoproduction from a large flux of high-energy photons}

\correspondingauthor{Matthew R. Mumpower}
\email{mumpower@lanl.gov}

\author[0000-0002-9950-9688]{Matthew R.\ Mumpower}
\affiliation{Theoretical Division, Los Alamos National Laboratory, Los Alamos, NM 87545, USA}
\affiliation{Center for Theoretical Astrophysics, Los Alamos National Laboratory, Los Alamos, NM 87545, USA}

\author[0000-0001-9891-0018]{Tsung-Shung H. Lee}
\affiliation{Physics Division, Argonne National Laboratory, Argonne, Illinois 60439, USA}

\author[0000-0003-1707-7998]{Nicole Lloyd-Ronning}
\affiliation{Center for Theoretical Astrophysics, Los Alamos National Laboratory, Los Alamos, NM 87545, USA}
\affiliation{Computational Division, Los Alamos National Laboratory, Los Alamos, NM 87545, USA}

\author[0000-0002-8825-0893]{Brandon L.\ Barker\textsuperscript{$\dagger$}}
\affiliation{Center for Theoretical Astrophysics, Los Alamos National Laboratory, Los Alamos, NM 87545, USA}
\affiliation{Computational Division, Los Alamos National Laboratory, Los Alamos, NM 87545, USA}

\author[0000-0002-7893-4183]{Axel Gross}
\affiliation{Theoretical Division, Los Alamos National Laboratory, Los Alamos, NM 87545, USA}
\affiliation{Center for Theoretical Astrophysics, Los Alamos National Laboratory, Los Alamos, NM 87545, USA}

\author[0000-0003-1758-8376]{Samuel Cupp}
\affiliation{Theoretical Division, Los Alamos National Laboratory, Los Alamos, NM 87545, USA}
\affiliation{Center for Theoretical Astrophysics, Los Alamos National Laboratory, Los Alamos, NM 87545, USA}

\author[0000-0001-6432-7860]{Jonah M.\ Miller}
\affiliation{Center for Theoretical Astrophysics, Los Alamos National Laboratory, Los Alamos, NM 87545, USA}
\affiliation{Computational Division, Los Alamos National Laboratory, Los Alamos, NM 87545, USA}

\renewcommand{\thefootnote}{}
\footnotetext{\textsuperscript{$\dagger$} Los Alamos Metropolis Fellow}
\renewcommand{\thefootnote}{\arabic{footnote}}

\begin{abstract}
We propose that neutrons may be generated in high-energy, high-flux photon environments via photo-induced reactions on pre-existing baryons. 
These photo-hadronic interactions are expected to occur in astrophysical jets and surrounding material. 
Historically, these reactions have been attributed to the production of high-energy cosmic rays and neutrinos. 
We estimate the photoproduction off of protons in the context of gamma-ray bursts, where it is expected there will be sufficient baryonic material that may be encompassing or entrained in the jet. 
We show that typical stellar baryonic material, even material completely devoid of neutrons, can become inundated with neutrons \textit{in situ} via hadronic photoproduction. 
Consequently, this mechanism provides a means for collapsars and other astrophysical sites containing substantial flux of high-energy photons to be favorable for neutron-capture nucleosynthesis. 

\end{abstract}



\section{Introduction}
\label{sec:intro}

The formation of the heaviest elements relies on astrophysical environments with a copious amount of neutrons \citep{Burbidge1957}. 
Neutrons can be found in the cosmos bound in atomic nuclei or in medium under extreme pressure. 
Free neutrons are rare, owing to a half-life of less than 15 minutes, see \cite{Wietfeldt2011, Serebrov2011} and references therein. 

In stellar interiors free neutrons are produced via low energy nuclear reactions, $\ce{^{13}C} + \alpha \rightarrow \ce{^{16}O} + n$, and $\ce{^{22}Ne} + \alpha \rightarrow \ce{^{25}Mg} + n$ \citep{Meyer1998, Maria2016, Wiescher2020}. 
Neutrons comprise the bulk of neutron stars, owing to the process of neutronization where the Fermi energy of electrons becomes high enough to energetically favor capture with protons via inverse beta decay \citep{Oppenheimer1939, Bethe1995}. 
An ample supply of neutrons make the merger of compact objects (neutron star-neutron star or neutron star-black hole) viable candidate sites for the rapid neutron capture ($r$ process) nucleosynthesis \citep{Lattimer1974, Freiburghaus1999, Rosswog2018}. 

Gamma-ray bursts (GRBs) may also be promising heavy element factories.  
As mentioned above, so-called short GRBs, `sGRBs', (GRBs with the prompt emission lasting less that  about two seconds) are believed to originate from the merger of two neutron stars \citep{Ab17} or a neutron star-black hole merger; the extremely neutron-rich material present in these merger events is conducive to a robust $r$ process. 
In the context of so-called long GRBs, `lGRBs', (GRBs with prompt gamma-ray emission lasting more than about two seconds), the picture is less clear. 
These GRBs are believed to originate from the collapse of a massive star \citep{Woos93, MW98} which forms an accretion disk engine around a compact object that launches a relativistic jet. 
Here we focus on a central black hole, in the so-called \textit{collapsar} model.\footnote{We note that a so-called \textit{magnetar} central engine with a magnetized neutron star compact object is also a viable model \citep{Usov92, DT92}, but we consider the black hole-disk central engine in this paper.} 
Several studies have suggested the presence of $r$ process nucleosynthesis in the disk around the newly born remnant black hole \citep{Fujimoto2007, Siegel19, Anand24}, while others have suggested that the $r$-process is severely suppressed in these systems \citep[e.g.][]{Miller2020, Just22, Blanch24}.

The presence of conditions suitable for nucleosynthesis in the GRB jet, cocoon (hot area encompassing the jet), and stellar envelope region remains relatively unexplored. 
Although the baryon density in GRB jets is necessarily very low in order to accelerate these jets to their inferred ultra-relativistic velocities, the baryon density in the stellar material surrounding a lGRB central engine, as well as the cocoon region created as the jet traverses this stellar envelope, have much higher baryon density. 
Additionally, dense shells ejected near the end of the massive star's life may provide yet another potential site for heavy element production, when the $\gamma$-rays from the jet interact with these regions. 

The interaction between light and matter in and around these regions can lead to extreme conditions that may be suitable for nucleosynthesis. High-energy, high-flux photon interactions with baryonic matter are relevant for nucleosynthesis in two ways. 
First, high energy photons have sufficient energy to break apart existing atomic nuclei via the process of photodisentigration. 
Second, for a range of high-energy photons, hadronic photo-interactions may transmute neutrons to protons or vice versa. 

In this work, we investigate a neutron production mechanism associated with a high flux of high-energy photons. 
We explore this possibility in the context of lGRBs associated with collapsars. 
In Section \ref{sec:picture} we provide an overview of the physical picture of lGRBs and how photohadronic processes can create neutrons relevant for nucleosynthesis. 
In the remaining sections, we provide the models and details which support this picture. 
Section \ref{sec:jet} covers the production of jet photons. 
Section \ref{sec:neutrons} covers the photoproduction of neutrons. 
Section \ref{sec:interactions} covers the dynamical interactions between regions of interest to the problem. 
We simulate nucleosynthesis based on our model parameters in Section \ref{sec:nucleosynthesis}. 
In Section \ref{sec:sites} we investigate other sites where photo-hadronic processes may emerge and assess their capacity for nucleosynthesis. 
We end with a discussion of potential observational signatures and concluding remarks. 

\section{Physical picture}
\label{sec:picture}

Gamma-ray burst jets are readily launched in black hole-disk systems (created when a massive star collapses) through the Blandford-Znajek (BZ) process \citep{BZ77, MT82}. 
Frame-dragging effects around a rapidly rotating black hole wind up magnetic fields (present in the disk, near the black hole horizon) and create a strong Poynting flux along the spin axis of the black hole \citep[for a brief discussion of the BZ process in the context of GRBs see][and references therein]{LR19}.  
 
As this energetic jet is launched, a ``fireball" forms---a radiation-dominated volume of plasma that is optically thick to pair production, typically containing a small amount of baryons. 
The fireball is accelerated, with Lorentz factor $\Gamma \propto r$, where $r$ is the distance of the jet head from the center of the star/central engine. 
For $10^{7}$ cm $\lesssim r \lesssim 10^{10}$ cm, the radiation energy in the fireball is converted to kinetic energy of the baryons. 
In the matter-dominated phase at $r \gtrsim 10^{10}$cm, the jet coasts with approximately constant ultra-relativistic velocity, with $\Gamma \gtrsim 100$. 
It may still be optically thick to pair production at this point, up until the opacity drops below the pair production threshold in the frame of the jet\footnote{We note that the optical depth to pair production is reduced by a factor of $\sim 1/\Gamma^{6}$ in the frame of the jet \cite{LS01}.}, which is expected to be at $r \gtrsim 10^{10}$cm. 

The flow is still optically thick to electron scattering until about $r \gtrsim 10^{12}$ cm, at which point the photons---particularly the $\gamma$-rays produced in internal dissipation processes such as shocks or magnetic reconnection events---can escape freely from the region. 
This is the so-called prompt emission phase of a GRB, at $10^{12}$ cm $ \lesssim r \lesssim 10^{15}$ cm, where synchrotron and inverse-Compton processes produce the initial highly variable ``burst'' of $\gamma$-rays, lasting tens of seconds and peaking at about 500 keV (in the observer's frame). 
Beyond this radius, the ``afterglow'' phase begins: this is the point where the front of the jet has swept up enough of the external medium such that the rest mass energy of the swept up material equals the kinetic energy of the jet, and the jet decelerates. \cite{Piran1999} contains a comprehensive summary of this general picture, including the relevant physical states and radii (see their Table 2) of the GRB jet. 
For additional reviews, see \cite{Pir04,ZM04, Mesz06, GRRF09, KZ15,Lev16}.

The jet has to travel through stellar material surrounding the central engine (material from the progenitor star that has not circularized into the disk around the black hole)---we term this stellar material the ``envelope''.
As it does so, it forms a cocoon---a hot, relatively dense (compared to the jet) region around the jet with a thermal X-ray spectrum \citep{RR02, NP18, DeColle18, DeColle2022}. 
In addition to these interactions in the envelope and cocoon regions, the jet photons can also interact with dense shells of matter ejected before the end of the progenitors lifetime. 

\begin{figure}[t]
  \centering
  \includegraphics[width=\columnwidth]{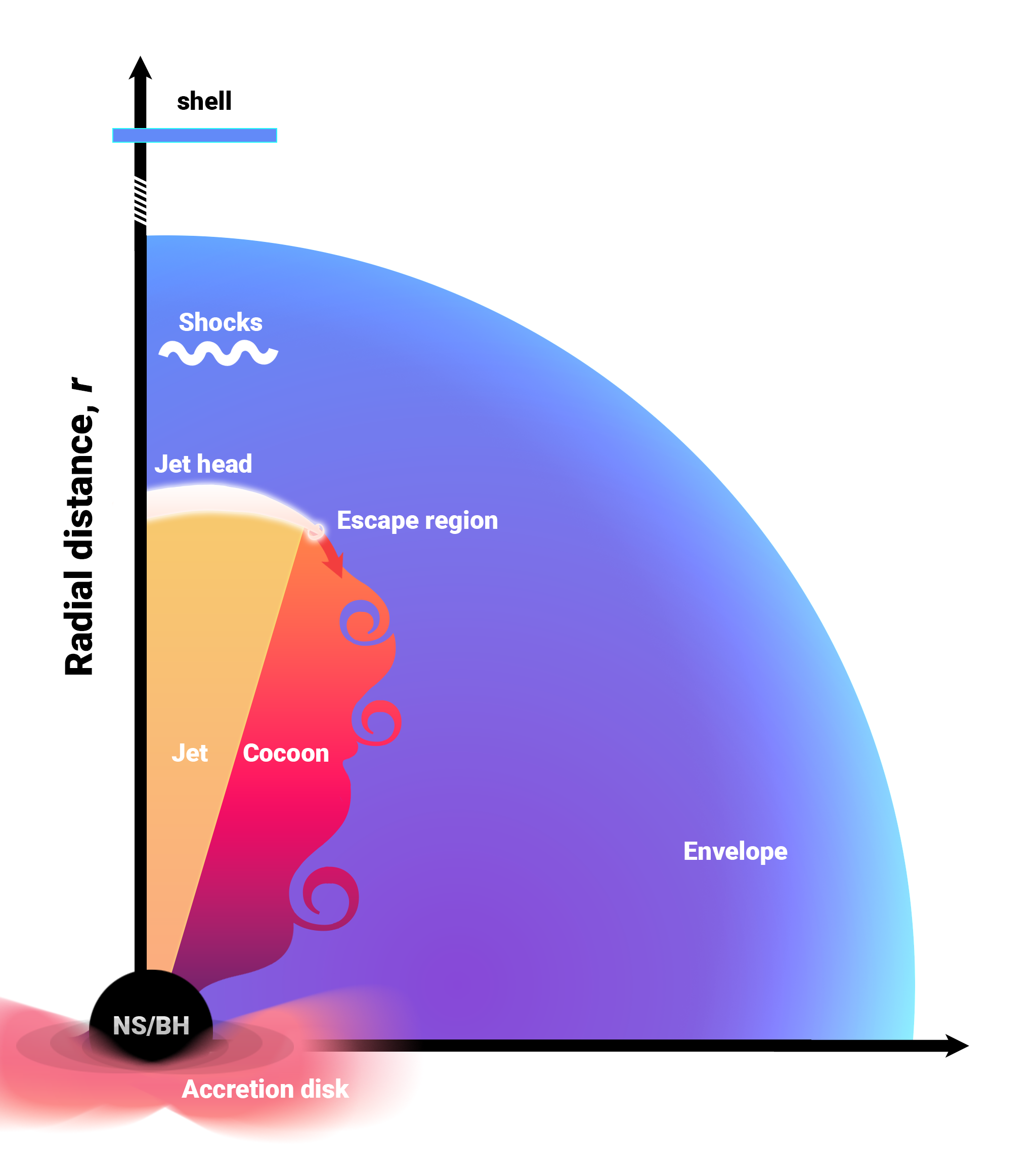}
  \caption{Schematic of the upper hemisphere of a GRB that highlights regions of interest. Each region is associated with different baryon and photon properties. Initially the jet head plows through existing stellar material in the envelope forming a cocoon. The jet may also encounter more dense regions or have bayronic loading which produces additional shocks. Once beyond the envelope, the jet may interact with ejected material in the shell region. With high photon flux and photon energy, the jet head could be a location for hadronic photoproduction. Subsequent nucleosynthesis could occur in the cocoon or shell regions; escaping high energy photons is key for forming heavy elements. Proportions shrunken or exaggerated for the purpose of visualization. }
\label{fig:cocoon}
\end{figure}

The regions of interest in a lGRB are shown in Figure \ref{fig:cocoon}. 
The jet head is the contact interface between the jet and the stellar envelope. 
It is in this region that existing baryonic material is exposed to a large flux of high-energy photons; the surrounding cocoon is therefore the most likely place for nucleosynthesis to develop. 

At the interface of light and matter in the jet head region, a large flux of high-energy photons enables exceedingly fast transmutation reactions between protons and neutrons while simultaneously producing mesons. 
These so-called photo-pion reactions have been traditionally associated with high-energy cosmic rays and neutrinos \citep{Berezinskii1975, Mannheim1995, Waxman1998, Murase2013}. 

Relativistic Lorentz contraction results in an effective compression of the stellar density as light interacts with matter; this process creates relatively high density (maximally: $2 \sqrt{2} \Gamma \rho_\textrm{env}$) in the jet head region as compared to the existing stellar density ($\rho_\textrm{env}$). 
Here $\Gamma$ is the Lorentz factor of the shock front in the frame of the star. 
In order for this higher density material to become active for nucleosynthesis, it must escape the jet head into the cocoon. 
Coincidentally, as the jet head plows through the star, it moves the transformed material out of the way at relativistic speeds. 

Relatively strong magnetic fields are expected to be generated in the jet environment \citep{Pir05, ZY11, HK13}\footnote{We note the very high magnetic field that launched the jet at the central engine site has decreased significantly at the radii where we suggest neutron production to emerge.} and will confine charged particles (protons and pions) along the axis of the jet with little chance of escape. 
Neutrons on the other hand have no charge, and therefore have a chance to escape; these particles are expected to escape at roughly the speed of the fluid velocity of the material that is being pushed out of the way of the jet. 
Conservation of baryon number reduces the baryon density in the jet head while the baryon density in the cocoon increases. 
The jet acts as its own sieve to filter out neutrons into the cocoon. 

Finally, the escape of a profusion of neutrons (at high density) mixing with relatively lower density material of the stellar envelope enables the cocoon region to become neutron rich. 
With a plethora of free neutrons, nucleosynthesis will proceed with haste. 
The efficiency with which baryons escape from the jet head plays a crucial role in determining the initial electron fraction ($Y_e$), which is a key parameter governing nucleosynthesis in the cocoon.
Furthermore, the continuous mixing of material within the cocoon affects the temporal evolution of both temperature and density. 
Variations in temperature and density can alter the synthesis of different isotopes thereby influencing the extent of nucleosynthesis.

\section{Jet photons}
\label{sec:jet}

In order to reach the bulk Lorentz factors inferred from observations\footnote{the variability of the prompt gamma-ray emission light curve combined with its non-thermal (optically thin) spectrum puts a constraint on the bulk Lorentz factor.}, which lie in the range of $10 \lesssim \Gamma \lesssim 1000$, the baryon number density in the jet is expected to be in the range $10^{-5} \ {\rm cm^{-3}} \lesssim {\rm n_{b}} \lesssim 10^{-3} \ {\rm cm^{-3}}$ \citep[e.g.][]{Pir04}. 
If baryonic loading of the jet is higher than this, the jet cannot be effectively accelerated and efficient production of high-energy photons is unlikely \citep{Pir04}. 
Hence, it is unlikely that there is significant hadronic interactions within the jet itself. 

\subsection{Functional form of the jet photon flux}

We model the photon flux of the jet head region using a jointly assembled power law. 
The functional form of the piecewise power law is
\begin{equation}
    \label{eqn:grb_flux}
    N(E_\gamma) = \begin{cases}
        A \left( \dfrac{E_\gamma}{E_{\text{pivot}}} \right)^\alpha \exp\left( - \dfrac{E}{E_0} \right), & \text{for } E_\gamma < E_{\text{break}} \\
        C \left( \dfrac{E_\gamma}{E_{\text{pivot}}} \right)^\beta, & \text{for } E_\gamma \geq E_{\text{break}}
          \end{cases}
\end{equation}
where $E_\gamma$ is the photon energy (keV), A  is a normalization constant to make the units ($\text{photons/cm}^2/\text{s}/\text{keV}$),  $E_{\text{pivot}}$ is the pivot energy (keV), $\alpha$ is the low-energy photon index, $\beta$ is the high-energy photon index and $E_0$ is the break energy (keV). 
The constant C is not independent, it is determined from a combination of the above parameters. 
The behavior of this power law at high photon energies depends on $\beta$. 
While this phenomenological formula is motivated from the fit to observed GRB spectra \citep{Band1993} and used here for its simplicity, more detailed studies show high-energy photon production is substantive in GRB environments \citep{Razzaque2004, Giannios2007, Asano2013}. 

\begin{figure}[t]
  \centering
  \includegraphics[width=\columnwidth]{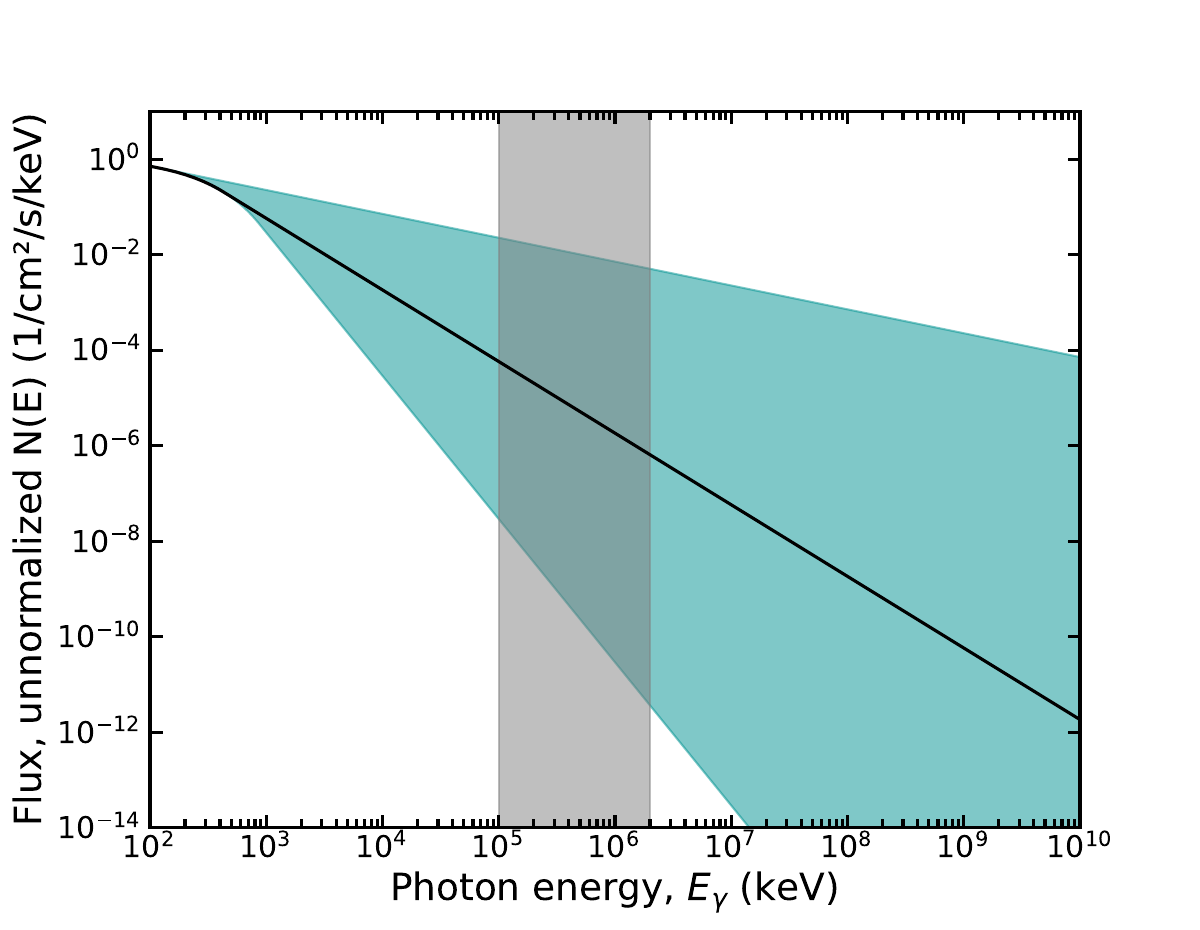}
  \caption{Schematic showing the range of relative decrease in photon flux as a function of photon energy depending on the parameter $\beta$; $\beta=-1.5$ shown in black. The grey hatched region is of interest for hadronic photoproduction. }
\label{fig:grb_flux}
\end{figure}

Figure~\ref{fig:grb_flux} shows the range of behavior for $\beta \in (-3.0, -0.5)$ with a normalization constant $A=1$, $\alpha=-0.1$, $E_\text{pivot} = 100.0$ (keV) and $E_0 = 300$ (keV). 
The region of interest for hadronic photoproduction is indicated by the grey hatched region and extends in photon energy approximately between $10^5$ keV to $2\times10^{6}$ keV. 

We use the normalization constant, $A$, to explore the photon flux in the regions of interest. 
When $A \neq 1$, we change notation $N(E_\gamma) \rightarrow \Phi(E_\gamma)$. 

\subsection{Physical (source) photon flux through the jet head}

We would like to estimate the $\gamma$-ray flux in the regions discussed in the previous section, particularly the flux impinging on the stellar envelope material at the jet head. 

For distances of $r \sim 10^{10}$ cm, the fireball is expected to still be in the optically thick regime. 
However, many studies \cite{Peer06, GS07, Bel10, Vurm11, Brom11a,Ahl19} have shown that the spectrum in the sub-photospheric region can undergo dissipation processes that create a pronounced high energy ``non-thermal-like'' $\gamma$-ray tail. 
These spectra can in principle be very hard, with power-law indices much harder than both a thermal spectrum and the typical $\beta \sim -2.5$ high energy index. 
Additionally, inverse Compton processes in the sub-photospheric region can lead to a copius number of even higher energy photons \citep{Ahl19}. 

If one assumes that the prompt $\gamma$-ray spectrum is made in this photospheric region (or, alternatively, that the jet becomes optically thin and creates non-thermal photons while still in the stellar envelope region), we can estimate the photon flux at the jet head interaction region. 
Drawing on Figure 2 of \cite{Ahl19}, the observed photon flux at $10^{6}$ keV is about $5 \times 10^{-3} \ {\rm ph/cm^{2}/s/keV}$.  
We can parameterize this $\gamma$-ray spectrum with a peak at $10$ keV at a normalization of $10 \ {\rm ph/cm^2/s/keV}$ with a high energy power-law photon index of $-1$.

In order to estimate the photon flux at the source, we need to multiply by a factor of $(d_{l}/r_{jh})^{2}$, where $d_{l}$ is the luminosity distance and $r_{jh}$ is the radius where the jet head is interacting with the envelope. 
If we assume a typical GRB distance of $d_{l} \approx 10^{28} {\rm cm}$ and the jet interaction radius of $r_{jh} \approx 10^{10}$ cm, we find the photon flux at the peak of the spectrum is $A = 10^{37} \ {\rm ph/cm^2/s/keV}$, whereas the flux at $10^{6}$ keV \citep[according to our extrapolation above from][]{Ahl19}, is on the order of $\Phi_{\gamma} \sim 10^{33} \ {\rm ph/cm^2/s/keV}$. 
We note that the jet head interacts with this envelope region over a timescale of 10’s to 100’s of seconds (from $10^{10}$ cm $< r < 10^{12}$ cm or even longer for more extended envelopes). 
The spectrum has energy injection from the jet so photon production should be sustained on this timescale.

\subsection{Additional high-energy photons from the cocoon}

We again note that the cocoon is expected to primarily emit a thermal spectrum, with a temperature that falls in the X-ray range \cite{RR02, Laz15, DeColle18}. 
However, \citep{DeColle18} pointed out that the thermal X-ray photons from the cocoon region can stream into the relativistic jet region and be inverse Compton (IC) scattered, potentially contributing to the high energy photons in the jet region. 
They calculate this IC spectrum, shown in their Figures 11 and 12 for two different scenarios of the jet initial conditions. 
If we assume a radius of $r_{IC} \approx 10^{13}$ cm where the cocoon photons interact with the energetic electrons in the jet, we estimate an IC photon flux of $\sim 10^{22} \ {\rm ph/cm^2/s/keV}$ at $10^{6}$ keV.

\subsection{Photon flux through previously ejected shells}

As the GRB jet collides with previously ejected shells, it is expected to be in its optically thin phase, possibly just at the time of onset of the afterglow. Therefore, both the prompt and afterglow high energy $\gamma$-ray flux will interact with the baryons in this shell. 

The photon flux in this region is expected to be $A \sim 10^{29} \ {\rm ph/cm^2/s/keV}$ at the peak of the spectrum ($\sim 500$ keV in the observer's frame). 
If we take a typical prompt emission power-law index of $\beta = -2$, the photon flux at $10^{6}$ keV (where the cross sections for hadronic photoproduction reaches a maximum) is roughly $\Phi_\gamma \sim 4 \times 10^{21} \ {\rm ph/cm^2/s/keV}$.

\section{Creating neutrons}
\label{sec:neutrons}

High energy photons on the (order of $10^{5}$-$10^{6}$ keV) incident on protons and neutrons produces hadrons (baryons and mesons). 
These processes are often referred to as photo-meson or photo-pion processes due to the production of pions. 
Dominant interactions include both direct and resonant (single pion) production, and multiple pion production \citep{Mucke1999}. 
Single meson production processes include
\begin{align}
    \label{p:gamma_1m_ncreate}  \gamma + p &\rightarrow \pi^{+} + n \ , \\ 
    \label{p:gamma_1m_nkill}    \gamma + n &\rightarrow \pi^{-} + p \ , \\ 
    \label{p:gamma_1m_nscatter} \gamma + n &\rightarrow \pi^{0} + n \ , \\
    \label{p:gamma_1m_pscatter} \gamma + p &\rightarrow \pi^{0} + p \ . 
\end{align}
The first process (\ref{p:gamma_1m_ncreate}) creates neutrons and the second process (\ref{p:gamma_1m_nkill}) destroys neutrons. 
The latter two processes are scattering processes that change the spectrum of the outgoing hadrons. 
We denote the cross section for neutron creation as $\sigma_{\gamma n}$, the cross section for neutron scattering as $\sigma_{\gamma n'}$, the cross section for proton creation as $\sigma_{\gamma p}$ and the cross section for proton scattering as $\sigma_{\gamma p'}$. 

Double meson production processes include 
\begin{align}
    \label{eqn:gamma_2m_1} \gamma + p &\rightarrow \pi^{0} + \pi^{0} + p \ , \\
    \label{eqn:gamma_2m_2} \gamma + p &\rightarrow \pi^{+} + \pi^{-} + p \ .
\end{align}
These processes may contribute to high energy neutrino production in astrophysical environments via the decay of the pions \citep{Mastichiadis2021}. 
Similarly, other meson production channels involving light pseudoscalar and vector mesons could contribute to neutrino production but are not considered here due to their short lifetimes. 
Henceforth we use the terms `photo-hadronic' or `hadronic photoproduction' to emphasize focus on the baryonic component of the products of the interactions that produce a single pion. 

\subsection{The ANL-Osaka model}
\label{sec:anlosaka}

The ANL-Osaka model provides a theoretical framework for understanding photoproduction processes involving hadrons \citep{Sato1996, Matsuyama2007, JuliaDiaz2007, Suzuki2010, Kamano2013, Kamano2019, Nakamura2018, Lee2019}. 
Developed collaboratively by researchers at Argonne National Laboratory (ANL) and Osaka University, this model integrates effective field theories with coupled-channel approaches \citep{Kamano2017b}. 
It furnishes a unified description of the world data (about 50,000 data points) of $\pi N,\ \gamma N \rightarrow MB$ with the meson-baryon channels $MB=
\pi N, \eta N, K\Lambda, K\Sigma$ and $\pi\pi N$ that have unstable $\pi\Delta$, $\sigma N$ and $\rho N$ states, where $N$ is a nucleon (proton or neutron) and $\Lambda$ and $\Sigma$ are baryons (B), and the other particles are mesons (M). 

The model provides a comprehensive description of the photoproduction mechanisms by incorporating multiple reaction channels and resonance contributions. 
It is based on the well-developed meson-exchange mechanisms and the assumption that the excitations of the nucleon to its excited states $N^*$ can be described by the $MB\rightarrow N^*$ vertex interactions. 
Within the Hamiltonian formulation \citep{Sato1996, Matsuyama2007}, the unitarity condition then requires that the scattering amplitudes $T_{MB,M'B'}(p_{MB},p_{M'B'};W)$ are defined by the following
coupled-channel equations:
\begin{eqnarray}
T_{\beta,\alpha}(p_\beta,p_\alpha;W)&=&V_{\beta,\alpha}(p_\beta,p_\alpha;W)\nonumber \\
&&+\sum_{\gamma}\int p^2,dpV_{\beta,\gamma}(p_\beta,p;W)\nonumber \\
&&\times G_\gamma(p;W)T_{\gamma,\alpha}(p,p_\alpha;W)
\label{eq:cc-eq}
\end{eqnarray}
with 
\begin{eqnarray}
V_{\beta,\alpha}(p_\beta,p_\alpha;W)=v_{\beta,\alpha}(p_\beta,p_\alpha)
+\sum_{N^*}\frac{\Gamma^\dagger_{N^*,\beta}(p_\beta)\Gamma_{N^*,\alpha}(p_\alpha)}
{W-M^0_{N^*}} \nonumber \\
&&
\label{eq:cc-v}
\end{eqnarray}
where $\alpha,\beta,\gamma=\pi N, \eta N, \pi\Delta, \sigma N, \rho N, K\Lambda, K\Sigma$; $G_\gamma(p;W)$ is the Green's function of the channel $\gamma$; $M^0_{N^*}$ is the bare mass of the excited nucleon state $N^*$; $v_{\beta,\alpha}(p_\beta,p_\alpha)$ is defined by meson-exchange mechanisms derived from effective Lagrangians, and the vertex
interaction $\Gamma_{N^*,\alpha}(p_\alpha)$ define $\alpha \rightarrow N^*$ transition. 

The results from the ANL-Osaka model are finalized in \cite{Kamano2013}. 
The predicted reaction amplitudes and the formula for using these amplitudes to calculate the cross sections of $\gamma N,\pi N \rightarrow \pi N, \eta N, K\Lambda, K\Sigma, \pi\pi N$ are given in \cite{Kamano2019}. 
The robust results are particularly adept at describing experimental data from photoproduction experiments, such as those conducted at Jefferson Lab and other high-energy facilities \citep{Matsuyama2007}. 
Theoretical predictions from the model aid in the interpretation of experimental results and the design of new experiments aimed at probing the excitation spectrum of nucleons \citep{Kamano2010} and offers insight into how hadronic properties evolve in different nuclear environments \citep{Wambach2003}. 

\begin{figure}[t]
  \centering
  \includegraphics[width=\columnwidth]{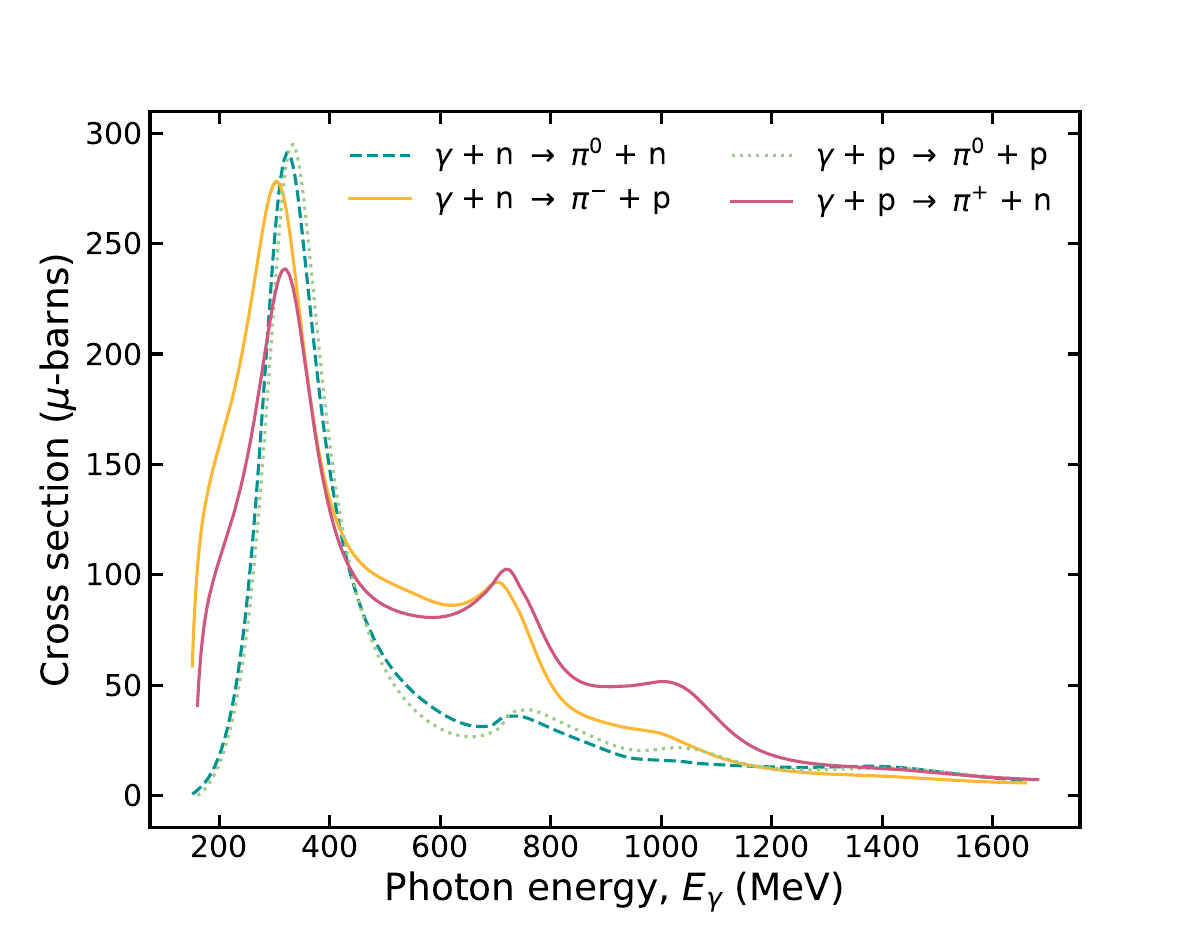}
  \caption{Hadronic photoproduction cross sections using the ANL-Osaka model. Solid lines indicate cross sections for neutron creation or destruction channels and broken lines indicate scattering cross sections. }
\label{fig:xs_gamma}
\end{figure}

In this work we use the aforementioned information to compute cross sections of the hadronic photoproduction processes (\ref{p:gamma_1m_ncreate}-\ref{p:gamma_1m_pscatter}) using the methodology outlined in \cite{Kamano2019}. 
The Appendix provides additional details. 

Cross section for the photoproduction processes (\ref{p:gamma_1m_ncreate}-\ref{p:gamma_1m_pscatter}) are shown in Figure \ref{fig:xs_gamma}. 
Transmutation cross sections between neutrons and protons are designated by solid lines while scattering cross sections are indicated by broken lines. 
The cross section for process (\ref{p:gamma_1m_ncreate}) matches well with experimental data with small uncertainty (not shown). 
The cross section for processes (\ref{p:gamma_1m_nkill} and \ref{p:gamma_1m_nscatter}) are purely theoretical predictions as neutron targets do not exist. 

All cross sections peak around $E_\gamma = 300$ MeV owing to the $\Delta(1232)$ resonance of nucleons \citep{PDG2022}. 
Proton creation from photons incident on neutrons dominates from threshold to approximately $E_\gamma \sim 700$ MeV. 
Higher energy photons, those with $E_\gamma \gtrsim 700$ MeV, favor neutron production. All cross sections tend towards zero above $E_\gamma \sim 1800$ MeV. 
The convolution of these cross sections with the environment's photon flux and baryonic density determine the reaction rate. 

\subsection{Photoproduction rates}

The creation rate of neutrons depends on the photon flux ($\Phi$), proton number density ($n_p$) and the cross section for process  (\ref{p:gamma_1m_ncreate}).
The neutron production rate is defined by
\begin{equation}
    \label{eqn:rn}
    r_\textrm{n} = n_\textrm{p} \int_{0}^{\infty} \Phi(E_\gamma) \sigma_{\gamma p}(E_\gamma) dE_\gamma = n_\textrm{p} \overline{r_p} \ ,
\end{equation}
where the integral runs over all possible photon energies. 
The units for Eq.~\ref{eqn:rn} are given in number of particles (neutrons) per unit volume per second. 
It is also convenient to to write this in terms of abundance, by dividing by the total baryonic density ($Y_x = \frac{n_x}{n_\textrm{total}}$), 
\begin{equation}
    \label{eqn:Rn}
    R_\textrm{n} = Y_\textrm{p} \int_{0}^{\infty} \Phi(E_\gamma) \sigma_{\gamma p}(E_\gamma) dE_\gamma = Y_\textrm{p} \overline{r_p}\ .
\end{equation}

Hadronic photoproduction of protons from neutrons (\ref{p:gamma_1m_nkill}) follows a similar form,
\begin{equation}
    \label{eqn:rp}
    r_\textrm{p} = n_\textrm{n} \int_{0}^{\infty} \Phi(E_\gamma) \sigma_{\gamma n}(E_\gamma) dE_\gamma = n_\textrm{n} \overline{r_n} \ ,
\end{equation}
and in terms of abundance, 
\begin{equation}
    \label{eqn:Rp}
    R_\textrm{p} = Y_\textrm{n} \int_{0}^{\infty} \Phi(E_\gamma) \sigma_{\gamma n}(E_\gamma) dE_\gamma = Y_\textrm{n} \overline{r_n}\ .
\end{equation}
Depending on the baryonic content of the material under consideration, one rate may dominate over the other, or a quasi-equilibrium may arise. 

\subsection{Inverse (meson collision) processes}

The hadronic processes (\ref{p:gamma_1m_ncreate}-\ref{p:gamma_1m_nkill}) which create or destroy neutrons also have inverse processes when a neutron or proton collides with a charged pion,
\begin{align}
    \label{p:pi_inv_p}  \pi^{+} + n &\rightarrow \gamma + p \ , \\ 
    \label{p:pi_inv_n}  \pi^{-} + p &\rightarrow \gamma + n \ . 
\end{align}
The short half-life of charged pions ($T_{1/2} \sim 10^{-8}$ s), and relatively slow interaction compared to forward processes limits these inverse processes. 
Below we show that the number of interactions is small compared to the forward rates. 

Consider a low energy ($ E_\pi < 100$ MeV) charged pion incident on a nucleon with kinetic energy $E_\pi = 20$ MeV that could arise from the aforementioned processes. 
The cross section for this reaction is on the order of $\sigma \sim 20$ (mb) \citep{Longo1962}. 
The mean free path is 
\begin{equation}
    \lambda = \frac{1}{n_b \sigma} \ ,
\end{equation}
where $n_b$ is the number density of the nucleons. 
The mean propagation distance before decay of the pion is 
\begin{equation}
    d = v \gamma T_{1/2} \ , 
\end{equation}
where $v$ is the pion velocity, $\gamma$ is the Lorentz factor, and $T_{1/2}$ the laboratory half-life. 
To estimate the number of pion interactions, take the ratio of these two quantities,
\begin{equation}
    n^{\pi}_\textrm{ints} = \frac{d}{\lambda} \ .
\end{equation}
The rest mass of a charged pion is approximately $140$ MeV. 
With $K=20$ MeV the Lorentz factor is $\gamma \sim 1.143$, which translates into a velocity of $v_\pi=0.484$c. 
The propagation distance before decay of the pion is then $d \sim 0.484 \times 1.143 \times 1.8 \times 10^{-8}$c $\approx 298.7$ cm. 
At low baryonic mass densities, $\rho_b \sim 1$ g/cm$^3$, the baryonic number density is $n_b = \frac{\rho_b}{M_n} \approx 6 \times 10^{23}$. 
The mean free path is $\lambda \sim \frac{1}{6 \times 10^{23} \times 2.0 \times 10^{-26}} \approx 8.3 \times 10^{2}$ cm. 
Therefore the number of interactions is $n^{\pi}_\textrm{ints} \sim \frac{298.7}{8.3 \times 10^{2}} \approx 3.56$. 

For higher energy pions the $\Delta$ resonance plays a role, increasing the cross section. 
For the sake of argument, a factor of 1000 increase in cross section is still not sufficient to compete with the forward processes,  (\ref{p:gamma_1m_ncreate}) and (\ref{p:gamma_1m_nkill}). 
If sufficiently large baryonic densities---for instance, those near nuclear saturation density---accompany the regions where hadronic photoproduction processes are operating, the inverse mesonic reactions will play a more substantial role. 
However, given the nature of the jet and stellar material considered here, we do not suspect that this type of density will be achievable. 
Hence, the inverse reactions to processes (\ref{p:gamma_1m_ncreate}) and (\ref{p:gamma_1m_nkill}) are unlikely to significantly contribute to the reaction dynamics. 

Consult the Appendix (Sec.~\ref{sec:spec}) for particle spectra associated with the forward reactions. 

\subsection{Hadronic photoproduction}
\label{sec:photoprod}

A complete picture of the spatial, energetic, and temporal statistics of particles requires use of the Boltzmann transport equations. 
These coupled integro-differential equations describe the evolution of the particle number as a function of phase space (position and momentum) and time. 
Because Boltzmann transport is extremely difficult to solve in practice, simplifying assumptions are made to make the problem more tractable. 
For instance, to follow the population of relativistic particles one can use the kinetic equation formulated by \cite{Ginzburg1964}. 

Assuming the distribution of particles is homogeneous in space (uniformly mixed) removes the spatial gradients in the Boltzmann description. 
With this simplifying assumption, one can imagine tracking the interactions in a small parcel of material that co-moves with a fluid element in the environment. 
The tracking of species without reference to spatial dependencies is often referred to as a `single zone' reaction network. 
Any interactions that inject or remove particles from this system can be treated as source or loss terms respectively in the simplified differential equation(s); this is a loss of information compared to the solution of Boltzmann transport. 
Contrastingly, multi-zone networks keep track of the movement of particles between neighboring fluid elements, see e.g. work by the NuGrid collaboration \citep{Herwig2008,Battino2019}, giving insight into spatial dependencies. 

A second simplifying assumption can be made regarding the energy distribution of the participating species. 
When particles follow relatively fixed energy distributions, each distribution may be integrated over to obtain the composition of the specified species as a function of a single independent variable, time. 
Reaction networks used in astrophysics that track elemental compositions make this assumption, typically with the distribution of species assumed to be Maxwell-Boltzmann like, see e.g.~\citep{Lippuner2017}. 

In what follows, we take both simplifying assumptions. 
The equations that describe composition simplify to
\begin{equation}
    \label{eqn:hadron_rxn_network}
    \frac{\partial n_x}{\partial t} = \sum_i C^{i}_x + \sum_j D^{j}_x + S_x + L_x \ ,
\end{equation}
where $n$ is the number density, the subscript $x$ indicates a hadron (baryon or meson) or lepton, $C^{i}_x$ represent interactions that create particle $x$, $D^{j}_x$ represents interactions that destroy particle $x$, $S_x$ indicates a source term that creates particle $x$ from `outside' the zone and $L_x$ is a loss term that removes particle $x$ from the zone. 

For relativistic particles, an entire suite of appropriate reactions should be considered; below we only consider several major species. 
Additionally, tracking the (nonthermal) distributions of the participating species is crucial, especially for estimating potential observable signals \citep{Guetta2004, Hummer2012}. 
Our aim here is to provide a first estimate of hadronic photoproduction by tracking composition alone. 
We plan to explore the many facets of an improved hadronic treatment in future work. 

\subsection{Interactions in the jet head}

Here we further simplify the hadronic reaction network by only considering transmutation reactions between protons, neutrons, and charged pions in the jet head and escaped neutrons outside the jet head region. 
This reduced set of possibilities provides an estimation for the interactions occurring at the jet head. 
The ordinary differential equations describing the population of these species are as follows
\begin{align}
    \label{eqn:hrxns1}
    \frac{\partial n_p}{\partial t} &= 
      n_\textrm{n} \overline{r_n} 
    + \frac{n_n}{\tau^n_\textrm{decay}} 
    - n_p \overline{r_p} 
    + n_n n_\pi v_\pi \sigma_{\pi} 
    - n_p n_\pi v_\pi \sigma_{\pi} \ , \\
    \label{eqn:hrxns2}
    \begin{split}
    \frac{\partial n_n}{\partial t} = 
      n_\textrm{p} \overline{r_p} 
    - \frac{n_n}{\tau^n_\textrm{decay}} 
    - \frac{n_n}{\tau^n_\textrm{esc}}
    - n_n \overline{r_n} \\
    - \ n_n n_\pi v_\pi \sigma_{\pi} 
    + n_p n_\pi v_\pi \sigma_{\pi}
    \end{split}
    \ , \\
    \label{eqn:hrxns3}
    \frac{\partial n^\textrm{esc}_n}{\partial t} &= 
      \frac{n_n}{\tau^n_\textrm{esc}} 
    - \frac{n^\textrm{esc}_n}{\tau^n_\textrm{decay}} \ , \\
    \label{eqn:hrxns4}
    \frac{\partial n_\pi}{\partial t} &=
      n_\textrm{p} \overline{r_p}
    + n_\textrm{n} \overline{r_n}
    - n_p n_\pi v_\pi \sigma_{\pi}
    - n_n n_\pi v_\pi \sigma_{\pi}
    - \frac{n_\pi}{\tau_\pi} \ ,
\end{align}
where $n_p$ is the proton number density in the jet head, $n_n$ is the neutron number density in the jet head, $n^\textrm{esc}_n$ are the neutrons that escape the jet head region corresponding to a timescale $\tau^n_\textrm{esc}$, $\tau^n_\textrm{decay}$ is the neutron decay lifetime, $n_\pi$ is the pion number density in the jet head, $v_\pi$ is the relative velocity of pions to nucleons and $\sigma_{\pi}$ is an estimation of the cross section of pions with nucleons, and $t$ is time. 

\subsection{Baryonic escape of the jet head material}

All charged particles are subject to the ``Hillas limit'' which refers to the theoretical maximum energy that charged particles can attain within an astrophysical accelerator, constrained by the size and magnetic field strength of the acceleration region \citep{Hillas1984, Blandford1987, Oka2024}. 
We argue below that the proton and charged pions do not escape the collimated jet due to the magnetic confinement of charged particles along the direction of the photon flux ($\tau^p_\textrm{esc} \approx \tau^\pi_\textrm{esc} \gg \tau^n_\textrm{esc}$). 
To see this, consider a charged particle in a magnetic field. 
One of the faster estimates for the escape time is Bohmian diffusion \citep{Bohm1949}
\begin{equation}
    \label{eqn:pbesc}
    \tau_\textrm{esc} \approx \frac{L^2}{D} \ , 
\end{equation}
where $L$ is the perpendicular distance that must be traveled to exit the jet and $D$ is the diffusion coefficient. 
The diffusion coefficient can be approximated by
\begin{equation}
    \label{eqn:bdiff}
    D \approx v \lambda \ ,
\end{equation}
where $v$ is the particle velocity and $\lambda$ is the mean free path. 
For particles spiraling in a magnetic field the mean free path is related to the gyro or Lamor radius, 
\begin{equation}
    \label{eqn:gyror}
    \lambda \approx \zeta r_L \ ,
\end{equation}
with the gyroradius, 
\begin{equation}
    \label{eqn:rl}
    r_L = \frac{\gamma m v}{q B} \ ,
\end{equation}
where $\zeta$ describes the turbulence of the magnetic field (effectively giving an uncertainty in the diffusion coefficient of a factor of a few \citep{Spitzer1960}), $\gamma$ is the Lorentz factor of the particle, $m$ is the particle mass, $q$ is the particle charge, and $B$ is the magnetic field strength. 
As discussed previously, we expect a significant magnetic field is generated and sustained in the fireball plasma with values $B \sim 10^{4} - 10^{10}$ G \citep{Pir05, ZY11, HK13}. 
 
The escape timescale for protons is then
\begin{equation}
    \label{eqn:taupesc}
    \tau^{p}_\textrm{esc} \approx \ \frac{q B L^2}{\zeta \gamma m_p v^2} \ .
\end{equation}
The perpendicular crossing distance is $L = r \ \textrm{tan}(\theta)$ where $\theta$ is the half opening angle of the jet and $r$ is the radial distance from the compact object, as in Figure~\ref{fig:cocoon}. 
Using fiducial numbers for the above quantities, a 300 MeV proton ($\gamma=1.32$) at $r=10^8$ cm with $\zeta = 1/3$, $q = 1.6 \times 10^{-19}$ C, $B = 10^{6}$ G, gives an escape time of $\tau^{p}_\textrm{esc} \sim 4.3 \times 10^{3}$ s. 

In contrast, the neutron escape timescale is estimated from calculating the average distance over the velocity of the fluid, $t^{n}_\textrm{esc} \approx \frac{L}{v}$. 
At a radius of $r=10^8$ cm, this is around $t^{n}_\textrm{esc} \sim 10^{-4}$ s. 
If the jet expands as it moves through the star, there will be larger perpendicular distances to cross further from the center, and the neutrons will have a longer escape time on average. 
If the jet remains collimated as it progresses, this timescale will remain relatively constant throughout the event. 

\subsection{A hadronic reaction network for the jet head}

As the jet is accelerated from $r \sim 10^7$ cm until about $r \sim 10^{10}$ cm, the jet head interacts with the surrounding stellar material, ``plowing'' the material, pushing it to the side and forming the hot cocoon region around the jet. 
High-energy photons from the jet region will interact with the shocked stellar envelope as this process proceeds, requiring a hadronic reaction network at the jet head. 
We note that the radius of a massive star is typically in the range $10^{10}$ cm $< R_{*} < 10^{12}$ cm, although gas can extend as far out as $10^{14}$ cm \citep{WH06}. 
The baryon number densities in the envelope are highly uncertain and---depending on stellar structure models---can be as low $n_{b} \sim 10^{15}$ cm$^{-3}$ \citep{Jiang23} or as high as $n_{p} \sim 10^{27}$ cm$^{-3}$ \citep[see, e.g., Figures 2 and 3 of][]{DeColle2022}. 

We model a time-dependent injection of photons, modifying the terms with $\overline{r_x}$ with a factor $\textrm{exp}(-\frac{t}{\tau_\textrm{inj}})$. $\tau_\textrm{inj}$ is bounded by either the time it takes for material to escape the jet head or by the time it takes for the shocked material behind the jet head to rarify. In both cases, this can be roughly estimated as the length of the jet head divided by the speed of light. We formally estimate the relevant length scale in Section \ref{subsec:dl:escape}; it is approximately $5\times 10^7$ cm, which gives a $\tau_\textrm{inj}$ of approximately $\tau_\textrm{inj} \approx 1.66\times 10^{-3}$ s.

For parameters of this model we use conservative values of $\tau^n_\textrm{decay} = 886$ s, $\tau^n_\textrm{esc} = 5 \times 10^{-4}$ s, $\tau_\pi = 10^{-8}$ s, and $\tau_\textrm{inj} = 10^{-4}$ s (larger values of $\tau_\textrm{inj}$ are more favorable, so this smaller value is conservative compared to our estimate above). 
The parameters of the photon flux are $A=10^{37}$ ph/cm$^2$/s/keV, $\alpha=-0.1$, $\beta=-1.5$, $E_\text{pivot} = 100.0$ (keV), and $E_0 = 300$ (keV). 

\begin{figure}[t]
  \centering
  \includegraphics[width=\columnwidth]{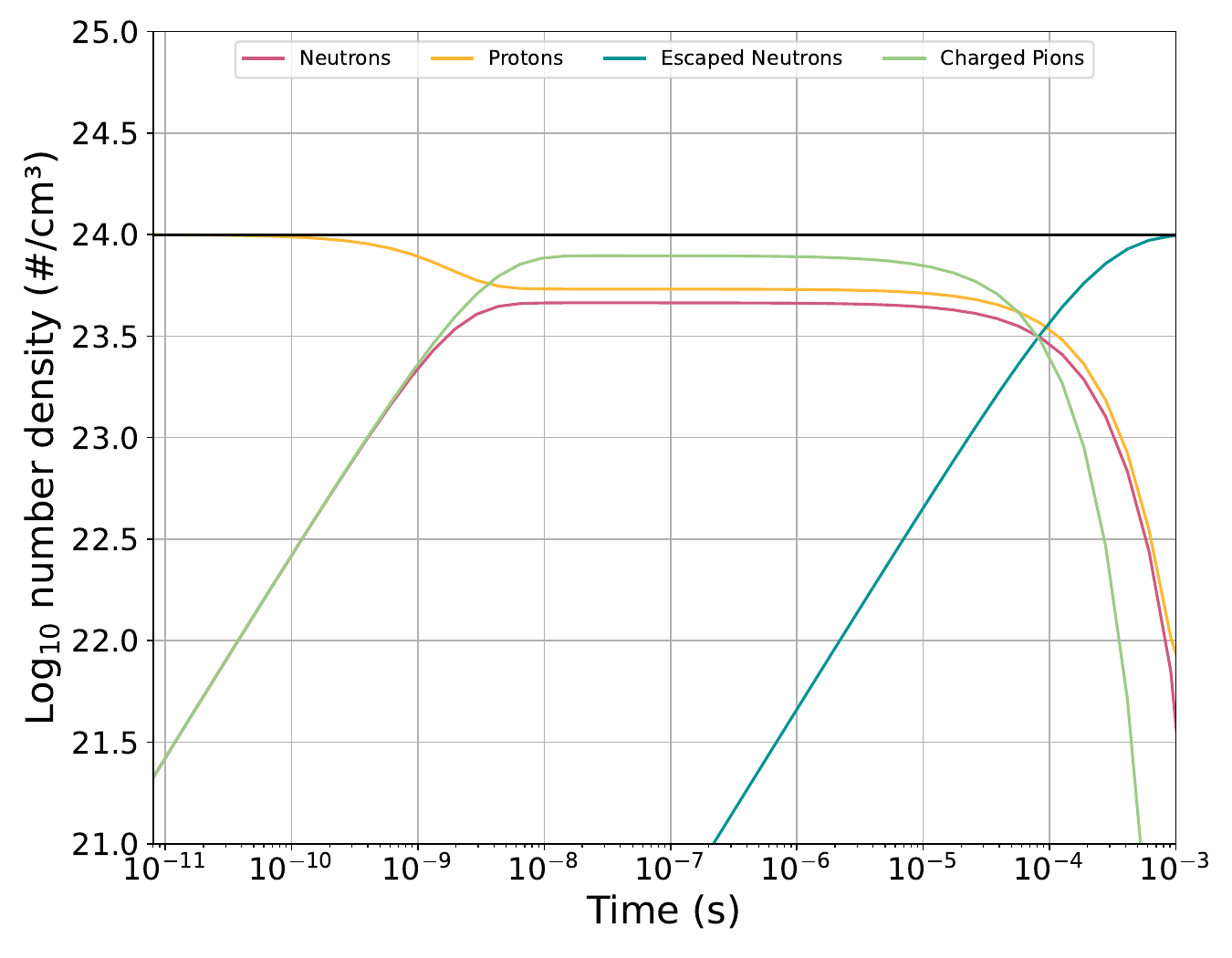}
  \caption{A simplified hadronic reaction network starting with 100\% protons (yellow). Neutrons (red) and pions (green) are produced nearly instantaneously. Baryon number is conserved (solid black line), despite neutrons escaping to another zone (blue); Meson number is not a conserved quantity (green).}
\label{fig:hrxn}
\end{figure}

Figure \ref{fig:hrxn} shows the temporal evolution of the hadronic network starting with a population of 100\% protons at $t=0$ s with a baryonic number density of $10^{24}$ cm$^{-3}$ or mass density of $\rho_b \sim 1$ g/cm$^3$. 
The baryon number (sum of the proton and neutron compositions) is conserved, as indicated by the solid black line. 
Given the high photon flux at $E_\gamma \sim 10^6$ keV, neutrons and pions are produced nearly instantaneously from the starting composition of protons. 
\textit{Let there be neutrons!}

The dynamics of hadronic interactions in the jet head depend on the aforementioned timescales. 
In this example we have assumed that the injection timescale is on the order of the escape time of the neutrons. 
If photon injection is very fast (decaying rapidly) relative to the escape time of neutrons, not all the baryons will transmute into neutrons and escape the jet head. 
We do not anticipate this to be a high probability scenario due to the continual injection of an extreme flux of photons in the jet head for long periods of time (10 to 100 seconds) as it plows through and escapes the envelope region. 

The above conclusions are independent of starting composition. 

If $\ce{^{4}He}$, or heavier nuclei (such as $\ce{^{12}C}$), instead comprise the initial composition, these nuclei will be photodissociated into free protons and neutrons as their binding energies are much less than the high-energy $\gamma$-rays. 
Under these starting assumptions, the only change to the dynamics is that the timescale for neutron production and transmutation is \textit{faster} owing to pre-existing (bound) neutrons. 

The lowest flux at $E_\gamma \sim 10^6$ keV suitable for considerable neutron escape is $\Phi_{\gamma} \sim 10^{25} \ {\rm ph/cm^2/s/keV}$. 
This is the minimal flux for an escape timescale of $t_\textrm{esc} \sim 10^{-4}$ s. 
Below this nominal value, the timescale for neutron production becomes longer than that of the neutron escape time limiting the possibility of neutron-rich nucleosynthesis. 
As the jet widens during its progression through the stellar material (recall Figure \ref{fig:cocoon}) it will lengthen the neutron escape time. 
Concurrently, the photon flux ($\Phi_{\gamma}$) is evolving from the interaction of the jet head with stellar material as well as from continual photon injection. 
It is therefore prudent for future simulations to study the balance of these two timescales. 

\subsection{Quasi-equilibrium}
\label{sec:quasi}

Figure \ref{fig:hrxn} shows a quasi-equilibrium between the neutron production (\ref{p:gamma_1m_ncreate}) and neutron destruction processes (\ref{p:gamma_1m_nkill}). 
We expect this to supervene in the jet head region with the duration dependent on the dynamics of the interaction between light and matter. 

Approximate equality between the rates (Eq.~\ref{eqn:Rn} and Eq.~\ref{eqn:Rp}) yields the ratio of neutrons to protons, 
\begin{equation}
    \label{eqn:quasi_eq}
    \frac{Y_\textrm{n}}{Y_\textrm{p}} \approx \frac{\int_{0}^{\infty} \Phi(E_\gamma) \sigma_{\gamma p}(E_\gamma) dE_\gamma}{\int_{0}^{\infty} \Phi(E_\gamma) \sigma_{\gamma n}(E_\gamma) dE_\gamma} \ .
\end{equation}

We estimate this ratio to be approximately between a range of values 0.5 to 0.85 using a range of steepness of the high energy $\gamma$ flux power law ($\beta$). 
With this range, the electron fraction, $Y_e$, takes values between 0.52 to 0.66 assuming the material only consists of free protons and neutrons. 
Relatively hard fluxes have smaller negative power laws, $\beta \gtrsim -1$, and support more symmetric matter. 
Steeper flux distributions, $\beta \lesssim -1$, favor more protons than neutrons. 

Nucleosynthesis with electron fractions in this range would be proton-rich. 
Assuming the photon flux dissipated, under near symmetric conditions the initial conditions generated from the quasi-equilibrium would result in elements like Ni ($Z=28$). 
In more extreme conditions, an overabundance of protons could yield some form of rapid proton capture nucleosynthesis \citep{Schatz2001}. 
For neutron-rich nucleosynthesis to happen, the neutrons must escape the jet head region as previously discussed. 

\section{Dynamical interactions between the jet, envelope, and cocoon, and between the jet and previously ejected shells}
\label{sec:interactions}

As the jet traverses the stellar envelope, it pushes material out of its way and a hot cocoon forms around the jet with temperatures falling in the $10$'s of keV range \citep{RR02, MLB10, Laz15}. 
The interface between the jet and the envelope creates a shock front and region between the jet and envelope. 
We have been referring to this region as the jet head.  
The shock interface can be described by the relativistic Rankine-Hugonoit jump conditions \citep{Taub1948}. 
In the frame of the star, the post-shock jet head baryonic density is Lorentz contracted. 
For an ultra-relativistic shock, the maximum value it can reach is
\begin{equation}
    \label{eqn:rh_rho_max}
    \rho^\textrm{max}_\textrm{head} = 2 \sqrt{2} \Gamma \rho_\textrm{env}
\end{equation}
where $\Gamma$ is the Lorentz factor of the jet (shock) and $\rho_\textrm{env}$ is the baryonic density of the (pre-shock) envelope \citep{Blandford1976}. 
In actuality, there will be some efficiency in which relativistic hydrodynamical processes increase the density of the jet head. 
We parameterize this efficiency via $\epsilon$, which ranges between 1 and $2 \sqrt{2} \Gamma$, 
\begin{equation}
    \label{eqn:rh_rho_eps}
    \rho_\textrm{head} = \epsilon \rho_\textrm{env} \ .
\end{equation}

\subsection{Functional form of the jet Lorentz factor}

The behavior of $\Gamma(r)$ for a lGRB can be written 
\begin{align}
    \label{eqn:gammar}
    \Gamma(r) =
    \begin{cases}
        \displaystyle \frac{r}{R_0} & \text{Acceleration Phase: } R_0 \leq r \leq R_s \\\\
        \displaystyle \eta & \text{Coasting Phase: } R_s \leq r \leq R_d \\\\
        \displaystyle \eta \left( \frac{r}{R_d} \right)^{-\frac{3}{2}} & \text{Deceleration Phase: } r \geq R_d \ ,
    \end{cases}
\end{align}
where $R_0$ is the starting radius of the jet, $R_s$ sets the coasting phase, $R_d$ sets the deceleration phase and $\eta$ represents the maximum of $\Gamma$ for a given burst. 
For continuity between the different phases, the coasting radius is constrained to $R_s = \eta R_0$. 

\subsection{Functional form of the stellar density}

Standard practice treats the baryonic mass density of the envelope region as a power law in radius which can be derived from a polytropic equation of state \citep{Woosley2002}. 
It has recently been shown by \cite{Halevi2023} that the baryonic density of the envelope region of a collapsing massive star can be written as 

\begin{equation}
    \label{eqn:rho_profile}
    \rho_\textrm{env}(r) = \rho_0 \left( \frac{r}{R_g} \right)^{-\delta} \left(1 - \frac{r}{R_{*}}\right)^3 \ ,
\end{equation}
where $r$ is the radial distance, $\rho_0$ is a scaling distance, $R_g$ is the gravitational radius of the compact object, and $R_{*}$ is the radius of the star. 
We take values of $\rho_0=2 \times 10^{9}$ g/cm$^3$, $\delta = 1.5$, $R_g=6.3 \times 10^5$ cm, and $R_{*}=10^{12}$ cm for these parameters. 
We are only interested in densities at $r > 10^{7}$ cm once the jet has formed, so values of the density $r < 10^{7}$ cm are not considered. 

\subsection{Interaction of the jet and envelope}

We set the jet parameters to $R_0 = 10^8$ cm (ten times launching point), $\eta = 100$ (maximum Lorentz factor), and $R_d = 10^{13}$ cm (interaction with previously ejected shell material). 

After the launch of the jet, relevant quantities are $\rho_\textrm{env}(r) \sim 3 \times 10^7$ g/cm$^3$ and $\Gamma(r) = 1$ at a distance of $r = 10^8$ cm. 
As the jet accelerates, it interacts with less dense material. 
At $r = 10^9$ cm, $\rho_\textrm{env}(r) \sim 3 \times 10^4$ g/cm$^3$ and $\Gamma(r) = 10$. 
At $r = 10^{10}$ cm, $\rho_\textrm{env}(r) \sim 9 \times 10^2$ g/cm$^3$ and $\Gamma(r) = 100$, entering the coasting phase. 

\begin{figure}[t]
  \centering
  \includegraphics[width=\columnwidth]{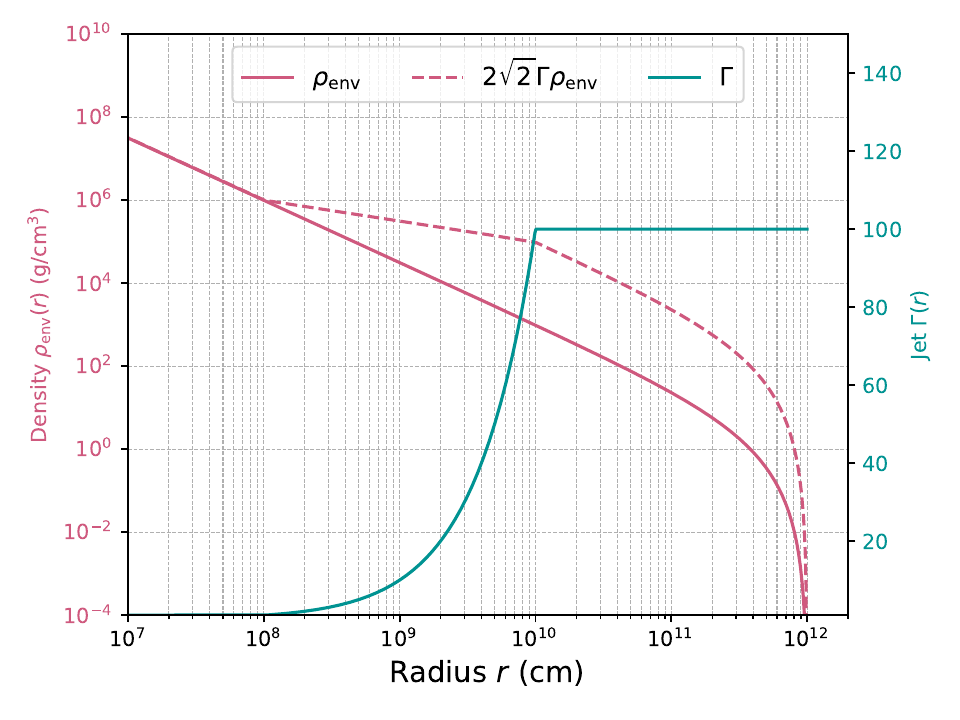}
  \caption{Density of the stellar envelope and jet Lorentz factor ($\Gamma$) as a function of radius. The effective density (dashed) is a product of these quantities. }
\label{fig:starnjet}
\end{figure}

Figure~\ref{fig:starnjet} shows the stellar envelope density and jet Lorentz factor as a function of radial distance. 
The product of these quantities yields the effective head density (recall Eqs.~\ref{eqn:rh_rho_max} and \ref{eqn:rh_rho_eps}). 
Depending on the dynamics, some of this high-density jet head material will escape and interact in the cocoon region. 
As a function of stellar radius, we therefore expect nucleosynthesis with a high starting baryon density to proceed at smaller radii and nucleosynthesis with low starting baryon density at larger radii. 

\subsection{Size of the escape region}
\label{subsec:dl:escape}
The pressure difference between the jet head and the surrounding envelope will create a funnel of material into the cocoon region. 
Subsequent interactions between the high density escaped material and cocoon material will be complex requiring hydrodynamical simulations to explore. 
Below we estimate the size of the escape region. 

The cross-sectional area through which the neutron rich material made in the jet head-envelope interaction region will be on the order of the thickness of the ``plowed'' or compressed gas in this region $\sim 10^{6}$ cm. 

To arrive at this estimate, consider the following.
An upper limit to the thickness of the ``plowed'' region is that its rest-mass energy density does not exceed the energy in the jet at that time---otherwise, the jet would decelerate and be choked. 
In other words, we have the condition that:
\begin{equation}
    \rho_{\rm head}(t) V_{\rm head} c^{2} < L_{j} \Delta t \ ,
\end{equation}
where $L_{j}$ is the power in the jet, $\Delta t$ is the timescale over which the jet head pushes on the plowed region and $V_{\rm head}$ is the volume of the plowed region.

The volume can be written as:
\begin{equation}
    V_{\rm head} = \pi r_{\rm cone}^{2} \Delta x = \pi (r_{jh} {\rm tan}(\theta))^{2} \Delta x \ ,
\end{equation}
where again $r_{\rm cone}$ is the radius of the conical opening at the jet head, $r_{jh}$ is the radius of the jet from the central engine, $\theta$ is the half-opening angle of the jet, and $\Delta(x)$ is the thickness of the plow region. 
We note that our assumption of a conical jet is conservative, and we expect that nucleosynthesis is even more likely for jets that are collimated by the surrounding stellar material.

Using $\rho_{\rm head}(t) = 2 \sqrt{2} \Gamma \rho_{\rm env}(t)$, we find that an upper limit to the thickness of this plowed region can be approximated as:

\begin{equation}
    \Delta x < \frac{L_{j} \Delta t}{2 \sqrt{2} \Gamma \rho_{\rm env}(t) \pi (r_{jh} {\rm tan}(\theta))^{2} c^{2}} \ ,
\end{equation}

\noindent or:

\begin{equation}
    \Delta x < 5 \times 10^{7} {\rm cm} \frac{(L_{j}/10^{50}erg s^{-1}) (\Delta t/10s)}{(\rho_{\rm env}(t)/10^{3}g cm^{-3})(r_{jh}/10^{10}cm)^{2}} \ ,
\end{equation}
where we have used a $\Gamma$ of 10, and a half opening angle of $5$ degrees. 

\subsection{Setting the initial electron fraction for nucleosynthesis}

The electron fraction of the material is difficult to ascertain due to its dependence on the dynamics. 
In the most extreme case, all baryons in the jet head are converted to neutrons, escape, and this neutron gas in turn mixes with material in the cocoon. 
The maximum ratio of the jet head number density (all neutrons) to the pre-existing stellar envelope density (all protons) is
\begin{equation}
    \label{eqn:yearg}
    \frac{n_\textrm{head}}{n_\textrm{env}} = \frac{2 \sqrt{2} \Gamma \rho_\textrm{env} m_p}{\rho_\textrm{env} m_n} = 2 \sqrt{2} \Gamma \frac{m_p}{m_n} = \frac{Y_n}{Y_p} .
\end{equation}
Considering only a population of neutrons and protons, the electron fraction is
\begin{equation}
    \label{eqn:yeratio}
    Y_e = \frac{Y_\textrm{p}}{Y_\textrm{n} + Y_\textrm{p}} = \frac{1}{Y_n/Y_p + 1} \ .
\end{equation}
The minimum electron fraction for a dense neutron gas streaming into a less dense gas of protons is then a function of $\Gamma$,
\begin{equation}
    \label{eqn:yenandp}
    Y^\textrm{min}_e = \frac{1}{2 \sqrt{2} \frac{m_p}{m_n} \Gamma + 1} \ .
\end{equation}
This value becomes the initial electron fraction of the cocoon. 
For $\Gamma = 1$, $Y^\textrm{min}_e \sim 0.26$ and a $\Gamma = 10$ yields $Y^\textrm{min}_e \sim 0.034$. 
If the stellar envelope instead consists of helium, the extreme case of total neutron conversion gives a minimum electron fraction of
\begin{equation}
    \label{eqn:yenandhe}
    Y^\textrm{min}_e \approx \frac{1}{4 \sqrt{2} (\frac{m_p}{m_n} +1)\Gamma + 1} \ .
\end{equation}
For $\Gamma = 1$, $Y^\textrm{min}_e \sim 0.081$ and for $\Gamma = 10$, $Y^\textrm{min}_e \sim 0.0088$. 
If achieved in nature, these excessively low electron fractions provide a natural explanation for the universality of the $r$-process pattern above $Z\gtrsim 50$ \citep{Cowan1999, Sakari2018, Farouqi2022}. 

For less extreme cases, some smaller fraction of neutrons will escape, shifting the electron fraction higher towards more equilibrium like conditions, $Y_e \sim 0.5$. 
The density of the escaped neutrons may also be reduced via dynamical processes. 
Therefore, depending on the complex dynamics that develops between the jet head, the stellar envelope and the cocoon, a wide range of electron fractions might be possible as the jet plows through the star. 
Using Eq.~\ref{eqn:rh_rho_eps} and starting from free protons, the initial electron fraction for nucleosynthesis under the assumption of perfect mixing and full escape of all neutrons is then
\begin{eqnarray}
    \label{eqn:Ye0}
    Y_e = \frac{1}{\epsilon \frac{m_p}{m_n} + 1} \ .
\end{eqnarray}

\subsection{Temporal evolution of the cocoon density for nucleosynthesis}

The evolution of density also plays a key role for nucleosynthesis as nuclear reactions depend on the square of this quantity \citep{Kippenhahn2012}. 
Expansion into free space under homologous ($r = v \times t$) assumptions yields a density evolution that depends on $t^{-3}$. 
Expansion of a shell of material depends on $t^{-2}$. 
Below we argue that the density emitted from the jet head first obeys a $t^{-1}$ evolution followed by a steeper evolution as the material expands. 

Consider a system consisting of two gases where the first gas is more dense than the second. 
As the first gas moves into the second, it picks up mass, 
\begin{equation}
    \label{eqn:dmdx}
    \frac{dm_1}{dx} = \rho_2 S \ ,
\end{equation}
where $m_1$ is the mass of the first gas, $\rho_2$ is the density of the second gas and $S$ is the cross sectional area traversed over distance $dx$.
As a function of distance, the mass of the first gas is
\begin{equation}
    \label{eqn:mx}
    m_1(x) = m_{1} + \rho_2 S x \ ,
\end{equation}
where $m_{1}$ is the initial mass of gas 1. 
Let the gases be mixed once $m_1(x=L) = 2 m_{1}$. 
The mixing length is 
\begin{equation}
    \label{eqn:mixdist}
    L = \frac{m_1}{\rho_2 S} \ .
\end{equation}
The mixing timescale is
\begin{equation}
    \label{eqn:mixtime}
    \tau_\textrm{mix} = \frac{L}{v} \ ,
\end{equation}
where the velocity, $v$ of the fluid is a substantial fraction of the speed of light; here we assume $v=0.8 c$. 
The mass density of gas 1 as a function of time is then
\begin{equation}
    \label{eqn:rhogas1}
    \rho_1(t) = \frac{\rho_1}{1 + \frac{v t}{L}} = \frac{\rho_1}{1 + \frac{t}{\tau_\textrm{mix}}} \ ,
\end{equation}
where $\rho_1$ is the initial density of gas 1. 

As the gas continues to mix it opens more degrees of freedom in spatial directions, on a different timescale leading to a steeper power law. 
The density evolution in time for nucleosynthesis can be modeled as 
\begin{equation}
    \label{eqn:rhot}
    \rho(t) = \rho_0 \left( 1 + \frac{t}{\tau_1} + \left(\frac{t}{\tau_2}\right)^\xi \right)^{-1} \ ,
\end{equation}
where $\xi > 1$. 

As an example, for a $\rho_2 = 4 \times 10^{4}$ g/cm$^3$, $m_1 = 10^{20}$ g, and $S = 10^{6}$ cm$^{2}$, the first timescale can be estimated to be around $\tau_1 = \tau_\textrm{mix} \sim 3.5 \times 10^{-2}$ s. 
In what follows we treat $\tau_1$, $\tau_2$, and $\xi$ as adjustable parameters. 

Because the cocoon is hot (relative to the stellar envelope), the density of gas in this region is expected to be smaller than in the jet head-envelope interaction region. 
The cocoon density may mix with the stellar envelope on relatively fast or slow timescales. 
For the latter case, simulations of the cocoon have shown an average mass density of $\rho_{c} \sim 1 \ {\rm g/cm^{3}}$ (or corresponding baryon number density $n_{b} \sim 10^{24}$ cm$^{-3}$) that remains relatively constant from about $r_{\rm cocoon} \sim 10^{9} - 10^{11}$ cm over timescales on the order of $\sim 10$'s to $100$'s of seconds \citep{Sal20, Suz22, DeColle2022, Gott22}. 
Beyond this timescale, the cocoon has expanded and its density drops off steeply \citep[see, e.g.][]{Suz22}. 
In the Section \ref{sec:nucleosynthesis} we consider the case of moderately and slowly expanding material using Eq.~\ref{eqn:rhot} in the context of nucleosynthesis. 

\subsection{Jet interaction with previously ejected shells}

Near the end of a massive star's life, the star can undergo episodic eruptions of shells of material \citep{smith06, Mes12, herwig14, fuller18, mauerhan18}. 
These shells can have masses that lie in the range between 0.1$M_{\odot}$ to a few $M_{\odot}$ \citep[e.g.,][]{Moriya2017, Morozova2018}. Depending on when in the massive star's lifecycle they were ejected and at what velocity, they are typically located anywhere from $r \sim 10^{13}$ cm - $10^{20}$ cm away from the center of the star. 
The widths of the shells typically range from $10^{13}$ cm $< R_{\rm shell} < 10^{15}$ cm.  

\begin{figure}[t]
  \centering
  \includegraphics[width=\columnwidth]{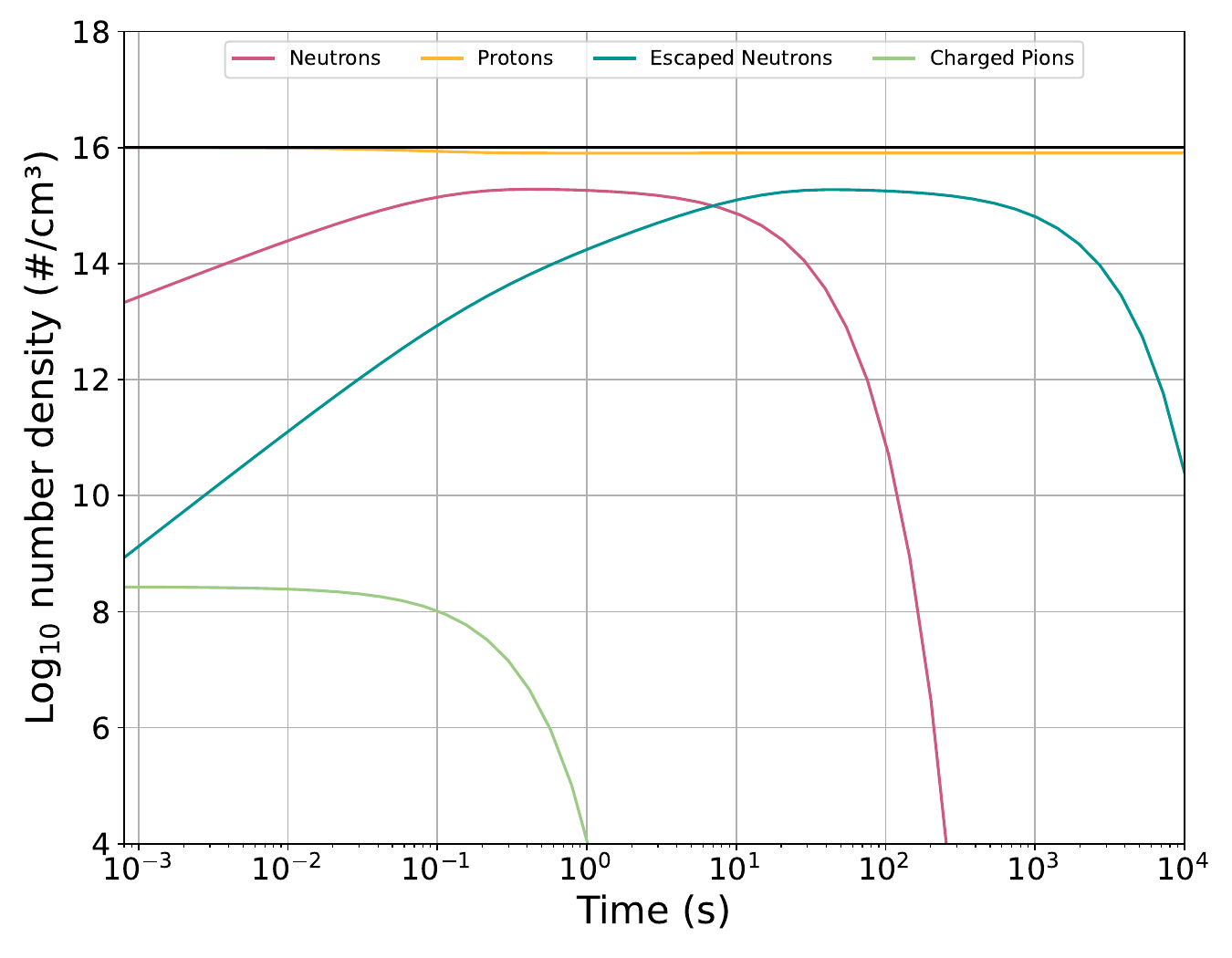}
  \caption{Hadronic reaction network showing the interaction between the jet and a shell of previously ejected material outside the start at $r = R_d$. Neutron production is less efficient in this region. }
\label{fig:hrxn_shell}
\end{figure}

If we consider a 1 $M_{\odot}$ shell at a distance of $r \sim 10^{15}$ cm with a width of $R_{\rm shell} \sim 10^{13}$ cm, this leads to a baryon number density of $n_{b} \sim 10^{14}$ cm$^{-3}$. 
While this is too diffuse to produce substantive nucleosynthesis, the production of free neutrons can still follow from photohadronic interactions in this region, albeit less efficiently than earlier in the life of the jet. 
At a distance of $r = R_d = 10^{13}$ cm, the baryon number density is roughly $n_{b} \sim 10^{16}$ cm$^{-3}$. 
The photon flux from the jet is reduced, $\Phi(E_\gamma=10^{6} \textrm{keV}) \sim 10^{21}$ ph/cm$^{2}$/s/keV, and the creation time for neutrons balloons to roughly a second. 
Similarly due to the wider opening area of the jet, the escape time for neutrons extends to $\tau^{n}_\textrm{esc} \sim 10$ s. 
The photon injection timescale was set to $\tau_\textrm{inj} = 0.1$ s. 
Figure \ref{fig:hrxn_shell} shows the behavior between the jet head and the previously ejected material. 
It should be noted that we do not distinguish between protons in the jet and those that escape the region in this calculation. 

\subsection{The case of a choked jet}

In the case of a choked jet, the burst does not have enough power to continually plow through the star to reach the outer envelope nor to previously ejected shells of material. 
This phenomena can also arise due to dynamical instabilities \citep{Bromberg2015, Gottlieb2020}. 
It is unclear how much nucleosynthesis could result, but it is likely to be  less than the case when the jet plows full steam ahead. 
Consequently, because of the `hidden' nature of such an event, nucleosynthesis could ensue without producing a measurable $\gamma$-ray signal to an outside observer. 
Instead, neutrinos may be used as potential messengers \citep{Meszaros2001,Murase2013b,Senno2016,Denton2018}. 

\section{Neutron-rich nucleosynthesis}
\label{sec:nucleosynthesis}

In this work we focus the discussion of nucleosynthesis to neutron-rich outcomes in the cocoon region where it is believed the bulk of the nucleosynthesis will transpire. 
Based off the discussion in previous sections, the electron fraction in this region can become exceedingly neutron-rich. 
A well resolved magnetohydrodynamical simulation for the distribution of $Y_e$ is the subject of further investigation. 

We simulate nucleosynthesis using version 1.6.0 of the Portable Routines for Integrated nucleoSynthesis Modeling (PRISM) reaction network \citep{Sprouse2021}. 
This network includes all relevant low-energy nuclear reactions \citep{Mumpower2024}, including those necessary to track nuclear fission \citep{Sprouse2020}. 
It should also be noted that the timescale for neutrons to thermalize from their high energy production ($E_n \gtrsim 300$ MeV) is thought to be fast due to the relatively large density of the spawning material in the jet head region. 
Therefore, neutron-induced reactions may be described by Maxwellian averaged cross sections that are functions of the local temperature, as is standard practice. 

The initial cocoon baryonic density is assumed to be of the form of Eq.~\ref{eqn:rhot}. 
For an adiabatically expanding gas the temperature is given by
\begin{equation}
    \label{eqn:tempfromrho}
    T(t) = T_0 \left( \frac{\rho(t)}{\rho_0} \right)^{\gamma-1} \ , 
\end{equation}
where $\gamma$ is the adiabatic index; here $\gamma=\frac{4}{3}$ for a radiation dominated gas. 

We model three interesting cases of outflow material in the cocoon region. 
For simulation (a) we follow inefficiently mixed material that expands like a shell at late times. 
The parameters are $T_0 = 2$ GK, $\xi=2$, and $r=10^9$ cm, with a pessimistic mixing $\epsilon=2.0$. 
The initial density is $\rho_0 = 3.2 \times 10^{4}$ g/cm$^3$, and timescales are set to $\tau_1 = \tau_2 = 3.5 \times 10^{-2}$ s with a starting $Y_e = 0.334$. 
For simulation (b) we take the same starting radius, except now we use the maximal mixing epsilon ($\epsilon = 2 \sqrt{2} \times 10 \sim 28.3$). 
The initial density is then $\rho_0 = 8.9 \times 10^5$ g/cm$^3$ and the starting $Y_e$ is $0.034$. 
The remainder of the parameters are the same. 
The higher density neutron region may interact in the cocoon region for some time. 
To explore this possibility we consider a third simulation (c), where we take $r=5.3 \times 10^{11}$ cm, $\rho_0 = 6.5 \times 10^3$ g/cm$^3$ ($\epsilon=282.8$), $T_0 = 0.1$ GK, $\tau_1 = 10^{-4}$ s, $\tau_2 = 10^{-1}$ s with a starting $Y_e = 0.0035$ and $\xi=3.5$. 

\begin{figure}[t]
  \centering
  \includegraphics[width=\columnwidth]{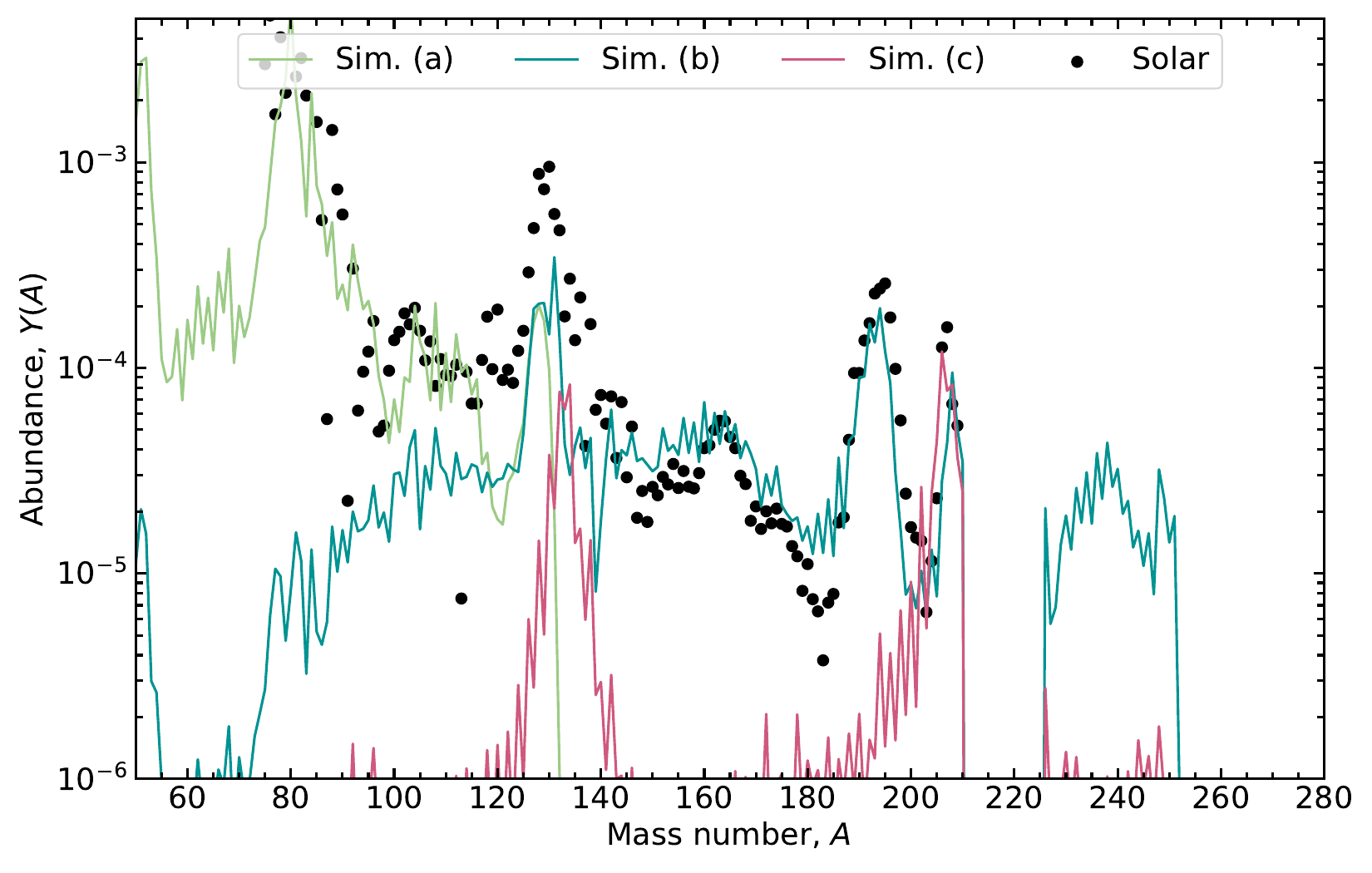}
  \caption{Resultant neutron-rich nucleosynthesis simulated with different condition sets (see text for details) at $4 \times 10^8$ seconds. Black dots indicate the solar $r$-process residuals. }
\label{fig:nucleo_nrich_sims}
\end{figure}

The resultant nucleosynthesis for the three cases is shown in Figure~\ref{fig:nucleo_nrich_sims}. 
Under scenario (a), weak $r$-process conditions are found creating nuclei up to the second $r$-process peak (mass number $A \sim 130$). 
Compare this outcome with simulation (b) --- the case of efficient mixing with all else equal to simulation (a) --- where a robust $r$-process is possible that produces a substantial amount of actinides. 
This simulation undergoes fission recycling producing a relatively flat pattern between $A=100$ and $A=170$. 
The second $r$-process peak at $A=130$ arises due to the $N=82$ shell closure. 

Under scenario (c), neutron capture is not as rapid, but persists for a longer time as compared to the previous cases. 
Photodissociation reactions do not play much of a role because the temperature starts and remains relatively low. 
Instead, there is a quasi-equilibrium between $\beta$-decay and neutron capture, much like a `cold' $r$-process of \cite{Wanajo2007}. 
Due to the long duration of neutron capture, the pattern looks markedly different from the solar residuals and the peaks are off-set to higher mass numbers. 
The behavior of nucleosynthesis in (c) is more like that of the $i$-process or intermediate neutron capture process than an $r$-process \citep{Cote2018, Choplin2021}. 
However, it is not a contemporary $i$-process believed to halt around the lead region. 
Instead we find that there is sufficient neutron capture to produce actinides, and even for nuclear fission to cycle some material back down to lighter atomic mass numbers. 
Intriguingly, these conditions reproduce with excellent agreement the lead peak of the solar residuals, as seen around mass number $A \sim 208$ in Figure~\ref{fig:nucleo_nrich_sims}. 
This is a fascinating scenario as lanthanides are produced which, once ejected, would result in a red kilonova that does not have solar-like proportions. 
It is also noteworthy that despite exceedingly neutron rich conditions, a full $r$-process does not occur. 
Rather, the density evolution in the cocoon is essential in understanding the resultant nucleosynthesis. 

The integration of all three scenarios generates an abundance pattern that closely aligns with the full range of solar $r$-process residuals, encompassing the first, second, third, and lead peaks. 
While alternative selections of low-energy nuclear models may induce localized deviations, the overall robustness of the combined pattern will remain consistent due to the astrophysical conditions regardless of the specific nuclear model employed \citep{Mumpower2016}. 
Finally, it is important to highlight that the nucleosynthesis processes considered in this study are largely insulated from neutrino interactions that could alter the electron fraction. 
This insulation is attributable to the lower densities present in the cocoon environment, which are significantly reduced compared to those typically encountered in explosive $r$-process conditions (e.g.~surrounding an accretion disk). 

\subsection{Ejection of material}

It is important to point out that any nucleosynthesis that occurs in the cocoon region is expected to be ``ejected'' or eventually gravitationally unbound from the central engine \citep{Laz15, Sal20}. 
Therefore the imprint of any nuclear processes that happens in this region will be present in the surrounding circumstellar region. 

When the jet deposits enough energy in the cocoon to roughly equal the binding energy of the star---$f L_{j}\Delta t \sim E_{\rm bind}$ where $f$ is the fraction of jet energy deposited into the cocoon, $L_{j}$ is the power in the jet, $\Delta t$ is the timescale for this limit to be reached, and $E_{bind} \sim GM^{2}_{*}/R_{*}$ is the binding energy of the star---the cocoon (and surrounding matter) will be unbound and able to travel into the circumstellar medium. 
At this stage, any elements made in the cocoon will be ejected into the interstellar medium.  
We estimate this timescale to be:
\begin{equation}
    \Delta t \approx 2 \times 10^{3} s  \frac{(M/20M_{\odot})^{2}}{(f/0.01)(R/10^{10}cm)(L_{j}/10^{50} erg/s)} \ .
\end{equation}

\noindent In other words, the cocoon should unbind on the order of about 2000 seconds for typical values of a GRB progenitor mass $M \sim 20 M_{\odot}$, radius of interaction $R \sim 10^{10}$ cm, and luminosity of the jet $L_{j} \sim 10^{50}$ erg/s, and if we assume roughly one percent of the jet energy is transferred to the cocoon.

To estimate the amount of material ejected from the cocoon region, consider the conical area plowed by the jet. 
The amount of mass is 
\begin{equation}
    \label{eqn:cone_mass_env}
    M = 2\pi (1 - \cos(\theta_0)) \int_{R_0}^{R_{*}} \rho_\textrm{env}(r) r^2 dr \ .
\end{equation}
Using previous values for these quantities we estimate that $M \sim 0.58$ \Msun between $R_0 = 10^8$ cm and $R_{*} = 10^{12}$ cm. 
Because the jet emits a bipolar distribution at both 0 and 180 degree angles a factor of two is needed, resulting in an estimate of 1.16 \Msun. 
If the jet starts ramping up at larger radii, less mass will be impacted. 
A lower bound for $R_0$ is $R_0 = 10^7$ cm in which case the total conical mass is $3.2$ \Msun; this value can be considered an upper bound. 
Therefore, somewhere between $0.1$ to $\sim $ few $M_{\odot}$, will become unbound from the central engine remnant. This estimate aligns with observational constraints on the cocoon ejected mass for the very special case where spectroscopic evidence of a cocoon was detected \citep{Izzo19}.
Even a small portion of this ejecta undergoing nucleosynthesis would be impactful for the chemical enrichment of galaxies. 

\section{Possible astrophysical sites}
\label{sec:sites}

Table~\ref{tab:astrosites} provides a list of other astrophysical phenomena with jets, where we might ask if the hadronic photoproduction processes discussed in this paper are possible.  We provide some fiducial numbers for the density in the jet regions, the expected high energy gamma-ray flux, and the relevant mass scale involved.  

The first line of the table gives the range of estimates included in this paper, at the site of collapsar GRB jets interacting with the cocoon and surrounding stellar material. 
The second line provides estimates in the context of jets from active galactic nuclei (AGN) interacting with the circumgalactic medium (CGM) \citep{Bland19}. 
The third fourth and fifth line give estimates for protostellar jet environments \citep{Koll18, Ray21}, pulsar wind nebulae \citep{Kar15}, and X-ray binary jets \citep{Fend04, Lew10}, respectively. 
We note that for objects like X-ray binaries and AGN jets, the regions around the jet---in contrast to GRBs---are generally very tenuous (the higher density regions are in the accretion disks themselves). 
The final line of the table gives partial estimates for jets from compact object (CO) mergers embedded in AGN disks. 
We note that because we do not have strong observational evidence for these objects \citep[although see the case of GRB191019A;][]{Laz23}, the photon flux from these jets at $10^{6}$ keV is not well constrained. 
And although the density in this case falls in a range where we do not expect strong $r$-process nucleosynthesis, it may be that in some cases the $i$-process is relevant if the photon flux is sufficient. 
We leave this line in the table as a site for potential future investigation.

We have not included in this table the traditional sites considered for r-process, like double neutron star \citep{Hoto18} or neutron star-black hole mergers \citep{Curt23} or accretion induced collapse of white dwarfs \citep{Cheong24}.
\begin{table*}
    \caption{Astrophysical Sites of Potential Interest for Hadronic Photoproduction}
    \label{tab:astrosites}
    \begin{tabular}{ccccc}
        \hline
         Site & $n_{b}$ (cm$^{-3})$ & $\Phi_{\gamma}$ at $10^{6}$ keV ($\rm ph/cm^{2}/s/keV)$ & nucleosynthesis (Y/N) & Ejected Mass $M_{\odot}$ \\
         \hline \hline
         
         GRB jet/stellar envelope & $10^{15} - 10^{27}$ & $10^{20} - 10^{33}$ & Y  & $0.1 \sim 3$ \\
         AGN jet/CGM environment & $10 - 10^{5}$ & $10^{10} - 10^{20}$ & N & na \\
         Protostellar jets & $10^{8} - 10^{20}$ & $10^{0} - 10^{3}$ & N & na \\
         Pulsar Wind Nebulae & $10 - 10^{4}$ & $10^{-3} - 10^{2}$ & N & na\\
         X-ray Binary Jets & $10 - 10^{4}$ & $10^{9} - 10^{12}$ & N & na \\
         CO mergers in AGN disks& $10^{15} - 10^{18}$ & ?? & --- & --- \\
         \hline
         \hline
         \vspace{1mm}
    \end{tabular}

\textbf{Note:} Fiducial numbers for baryon number density and photon flux at $10^{6}$ keV in different astrophysical environments with jets, where high energy gamma-rays may be produced.  The photon flux $\Phi_{\gamma}$ is estimated {\em at the source}. Ejected mass is listed as ``na" when significant heavy element nucleosynthesis is not expected.  Note this table does not include traditionally considered r-process sites like double neutron star mergers \cite{Hoto18} or accretion induced collapse of white dwarfs \citep{Cheong24}.

\end{table*}

\section{Potential observational signatures}
\label{sec:observations}

Here we list and discuss possible observable signatures of neutron production associated with GRBs.

\begin{itemize}
    \item If neutron creation is robust, and there is sufficient density in surrounding material, lGRBs may be associated with $r$-process nucleosynthesis, see e.g.~ \cite{Rastinejad2022, Yang2024}. Subsequent interaction between the radioactively decaying material and emitted light can produce a kilonova \citep{Metzger2019}. An extended duration of kilonova afterglow, particularly in the red spectrum, would indicate the presence of actinides \citep{Zhu2021, Barnes2021}; such a scenario has not yet been observed. 
    \item Alternatively, $\gamma$-ray lines stemming from nuclear transitions can also be used to diagnose the production of heavy elements \citep{Korobkin2020}. A line which would indicate the production of gold is the $E_\gamma=2.6$ MeV line associated with the $\beta$-decay of $\ce{^{208}Tl}$. Depending on the observational timescale of this line, it can be associated with a complete $r$ process that produces actinides \citep{Vassh2024}. Other characteristic nuclear lines exist---for example, neutron capture on protons to create deuterium, $n + p \rightarrow d + \gamma$ could lead to signature $\gamma$-ray emission with $E_\gamma=2.223$ MeV. Observable gamma-ray lines may be challenging to detect due to broadening effects, which arise from the Doppler shift caused by the relative velocity of the ejected material with respect to the observer. 
    \item Robust conditions that produce substantial nuclear fission are found to be plausible, in accordance with recent observational hints of fission in the cosmos \citep{Roederer2023}. As a result, signatures of fission may also be associated with GRBs in future observations \citep{Wang2020}. The production of $\ce{^{254}Cf}$, or other relatively long-lived heavy species undergoing fission may also provide a smoking gun signature of a complete $r$-process; this signature may be observable with the James Webb Space Telescope \citep{Zhu2018}. 
    \item If neutron creation is moderate coupled with a slowly evolving density, GRBs can be associated with an intermediate ($i$-process) nucleosynthesis. A full $i$-process may reach Pb (or beyond), yielding a signature of $\ce{^{208}Tl}$ \citep{Vassh2024}. As shown here, this process also produces lanthanides, making it difficult to distinguish between kilonova with a red spectral component that has been traditionally associated with the $r$-process. 
    \item If sufficient neutrons are created and sustained along the front of the jet head, or if there is sufficient neutron production as the jet interacts with previously ejected shell material, a neutron precursor event \citep{Metzger2014} could be associated with GRBs. 
    \item Neutron-deficient conditions (not explored here), those in which the proton production channel (\ref{p:gamma_1m_nkill}) is favored or sustained in some way, may yield the production of elements like $\ce{^{56}Ni}$ that have distinguishable light curves \citep{Colgate1980}.
    \item A high-energy `pion bump' could be associated with GRB spectra if relevant photo-hadronic processes occur in an optically thin region that escapes further processing \citep{Yang2018}. 
    \item High-energy pions will decay to produce high-energy neutrinos, which are easier to detect than low-energy neutrinos \citep{Valera2022}. Future neutrino detections connected with GRBs will provide a telltale signature of the proposed hadronic interactions and therefore provide a multimessenger signal for constraining heavy element formation. The simultaneous multi-messenger detection of photons and neutrinos offers a test of the weak equivalence principle (which assumes equality of gravitational and inertial mass), a foundational assumption of Einstein's General Relativitiy \citep{Waxman1997}.
\end{itemize}

\section{Conclusion}

We have estimated the production of protons and neutrons via photo-hadronic interactions from a large flux of high-energy photons. 
The production of neutrons via the process, $\gamma + p \rightarrow \pi^{+} + n$, is pertinent to the astrophysical origin of the heavy elements. 
Rather relying on pre-existing neutrons this physical mechanism produces neutrons \textit{rapidly} and \textit{in situ}. 
For this process to be relevant for nucleosynthesis, baryonic material must be present surrounding an astrophysical jet and the jet must contain charged particles on a longer timescale than it does neutrons. 
These two conditions provide strict limits on viable locations in the universe. 

As there is a wide range of GRBs, from short to long, one can naturally postulate that there will also be a range of resultant nucleosynthesis processes that can be associated with these jets. 
Long GRBs are of particular interest as they plow through a mass of dense stellar material for extended durations allowing ample time for photo-hadronic processes to prevail. 

We have explored a few interesting cases of neutron-rich nucleosynthesis in this work.  
Based on the initial population of protons or He in the stellar envelope, we believe that the cocoon is a promising region for neutron-rich nucleosynthesis. 
We have considered that nucleosynthesis in this region is capable of creating a complete rapid neutron capture ($r$-process) pattern as well as producing conditions viable for slower neutron capture, akin to the intermediate capture process ($i$-process). 
The final abundance patterns that arise from these cases are drastically different and warrant further investigation. 

Multi-scale, multi-physics modeling is required to accurately model astrophysical transients. 
Our work reinforces this perspective and provides a new incentive to add medium / high energy phenomenon to the modeling of these complex and interesting environments.

\newpage

\onecolumngrid
\section*{Appendix: ANL-Osaka model of hadronic spectra}
\label{sec:spec}

The ANL-Osaka model generates  the differential cross sections in the center of mass (CM) frame of the initial $\gamma N$ and the final $\pi N$ states.
In this notation, $N$ represents a nucleon (proton or neutron) and $\pi$ represents a charged or neutral pion. 
For the process $\gamma({\bf q}_c)+ N(-{\bf q}_c)\rightarrow \pi ({\bf k}_c) + N (-{\bf k}_c)$, the differential cross section is of the following form (omitting the spin and isospin indices)
\begin{eqnarray}
\frac{d\sigma_c}{d\Omega_c}(W,\theta_c)(W)=\frac{(4\pi)^2}{{\bf q}_c^2}
\rho_{\pi N}(W)\rho_{\gamma N}(W)
|<{\bf k}_c) |T_{\pi N,\gamma N}(W,\theta_c)|{\bf q}_c>|^2
\label{eq:dcrstcm}
\end{eqnarray}
where $\theta_c$ is the scattering angle defined by $cos\theta=\hat{{\bf k}_c}\cdot \hat {{\bf q}_c}$, the amplitude $<{\bf k}_c |T_{\pi N,\gamma N}(W,\theta)|{\bf q}_c>$ is calculated from the multiple amplitudes $E_{L^{\pm}}(W)$ and $M_{L^{\pm}}(W)$ listed on the
ANL-Osaka Website \citep{Lee2019}, and the invariant mass $W$ and the phase space factors are
\begin{eqnarray}
W&=&|{\bf q}_c|+E_N({\bf{q}_c})=E_\pi({\bf k}_c)+E_N({\bf k}_c)\,,\\
\rho_{\pi N}(W) &=&\pi\frac{|{\bf k}_c| E_\pi({\bf k}_c) E_N({\bf k}_c)}{W}\,,\\
\rho_{\gamma N}(W) &=&\pi\frac{|{\bf q}_c |^2|E_N({\bf q}_c)}{W}\,.
\end{eqnarray}
Here the energy is $E_a({\bf p})=\sqrt{{\bf p}^2+m_a^2}$ for
a particle $a$ with mass $m_a$ and momentum ${\bf p}$.

In the Laboratory (Lab) frame, the initial nucleon has momentum ${\bf p}_t$ and the scattering angles for the outgoing pion momentum ${\bf k}$ and the nucleon momentum ${\bf p}$ are defined by the energy and momentum conservations:
\begin{eqnarray}
|{\bf q}|+E_N({\bf p}_t) &=& E_\pi({\bf k})+E_N({\bf p})\label{eq:econ} \ ,\\
{\bf q}+{\bf p}_T&=&{\bf p}+{\bf k} \ , \label{eq:pcon}
\end{eqnarray}
The differential cross section in this Laboratory frame can be calculated from the CM differential cross section Eq.~({\ref{eq:dcrstcm}) by finding the invariant mass $W$ of the initial $\gamma N$ system 
\begin{eqnarray}
W&=&
[(|{\bf q}|+E_N({\bf p}_T))^2- ({\bf q}+{\bf p}_T)^2]^{1/2} \ ,
\end{eqnarray}
and by using the following transformation
\begin{eqnarray}
\frac{d\sigma_L}{d\Omega_k}(W,\theta_k)
=|\frac{\partial \Omega_c}{\partial\Omega_k}|
\frac{d\sigma_c}{d\Omega_c}(W,\theta_c) \ ,
\label{eq:dcrstlab-k}
\end{eqnarray}
where $\theta_k$ is the scattering angle of the outgoing pion with respect to
the incident photon momentum ${\bf q}$. Alternatively, we can also define the differential cross section in terms of the scattering angle $\theta_p$ of the outgoing nucleon momentum ${\bf p}$:
\begin{eqnarray}
\frac{d\sigma_L}{d\Omega_p}(W,\theta_p)
=|\frac{\partial \Omega_c}{\partial\Omega_p}|
\frac{d\sigma_c}{d\Omega_c}(W,\theta_c) \ .
\label{eq:dcrstlab-p}
\end{eqnarray}

In Eqs.~(\ref{eq:dcrstlab-k}) and (\ref{eq:dcrstlab-p}) the angles $\theta_k$ and $\theta_p$ are not independent because of the energy and momentum conservation conditions Eqs.(\ref{eq:econ})-(\ref{eq:pcon}). 
They are determined by the Lab momenta ${\bf k}$ (${\bf p}$) which can be calculated from the CM momentum ${\bf k}_c$ ($-{\bf k}_c$) by using the Lorentz Boost transformation:
\begin{eqnarray}
{\bf k}=B^{-1}(\vec{\beta})[+{\bf k}_c] \ , \label{eq:boost-k}\\
{\bf p}=B^{-1}(\vec{\beta})[-{\bf k}_c] \, \label{eq:boost-p}
\end{eqnarray}
where $\vec{\beta}$ is the velocity of the initial $\gamma N$ system:
\begin{eqnarray}
\vec{\beta}=\frac{{\bf q}+{\bf p}_T}{|{\bf q}|+ E_N({\bf p}_T)} \ .
\end{eqnarray}

By considering the Lorentz invariant condition and assuming that photons are in $Z$-direction, we then have
\begin{eqnarray}
(q-k)^2=(q_c-k_c)^2 \ ,
\end{eqnarray}
which leads to  
\begin{eqnarray}
|{\bf q}|E_N({\bf k})-|{\bf q}||{\bf k}|cos\theta_k 
=|{\bf q}_c|E_N({\bf c})-|{\bf q}_c||{\bf k}_c|cos\theta_c \ ,
\end{eqnarray} 
and hence 
\begin{eqnarray}
|\frac{\partial \Omega_c}{\partial\Omega_k}|=\frac{|{\bf q}||{\bf k}|}
{|{\bf q}_c||{\bf k}_c|} \ , 
\end{eqnarray}
for calculating the differential cross sections using Eq.~(\ref{eq:dcrstlab-k})
Similarly, we have
\begin{eqnarray}
|\frac{\partial \Omega_c}{\partial\Omega_p}|=\frac{|{\bf q}||{\bf p}|}
{|{\bf q}_c||{\bf k}_c|} \ ,
\end{eqnarray}
for calculating the differential cross sections using Eq.~(\ref{eq:dcrstlab-p}).

For a range of photon energies we are interested in, we can use the ANL-Osaka model to generate
$\frac{d\sigma_L}{d\Omega_k}(W,\theta_k)$ 
($\frac{d\sigma_L}{d\Omega_p}(W,\theta_p)$) in a range of $W$. 
We then can get pion (nucleon) spectrum for each $\theta_k$ ($\theta_p$) as a function of pion (nucleon) momentum $k$
($p$). 
Another information crucial to this problem is the outgoing particle spectrum in scattering angles $\theta_k$ or 
$\theta_p$ for a given incoming photon momentum ${\bf q}$.

It will be complicated to use the spectra generated from the above procedures. 
As a start, we may just need to see the spectrum of the total pions (nucleons)
with an averaged pion momentum $\overline{k}$ ($\overline{p}$) defined by
\begin{eqnarray}
\overline{k}^2&=&\frac{\int {\bf k}^2\frac{d\sigma_c}{d\Omega_c}(W,\theta_c) d\Omega_c}
{\int \frac{d\sigma_c}{d\Omega_c}(W,\theta_c) d\Omega_c}\label{eq:ave-k} \ ,\\
\overline{p}^2&=&\frac{\int {\bf p}^2\frac{d\sigma_c}{d\Omega_c}(W,\theta_c) d\Omega_c}
{\int \frac{d\sigma_c}{d\Omega_c}(W,\theta_c) d\Omega_c} \ , \label{eq:ave-p}
\end{eqnarray}
where ${\bf k}$ (${\bf p}$) can be calculated from the CM momentum ${\bf k}_c$ ($-{\bf k}_c$) by using Eqs.~(\ref{eq:boost-k})-(\ref{eq:boost-p}).

As a start, we perform calculations for the case that the initial nucleon
in the Lab frame is at rest with ${\bf p}_T=0$. 
The calculated total cross sections from threshold to about 1500 MeV of the photon momentum $q$ are shown in
Fig.~\ref{fig:xs_gamma}. 
The outgoing spectra corresponding to the averaged momenta are shown for the four cross sections of interest in Figure \ref{fig:specs}. 

\begin{figure}[htbp]
    \centering
    \includegraphics[width=0.45\textwidth]{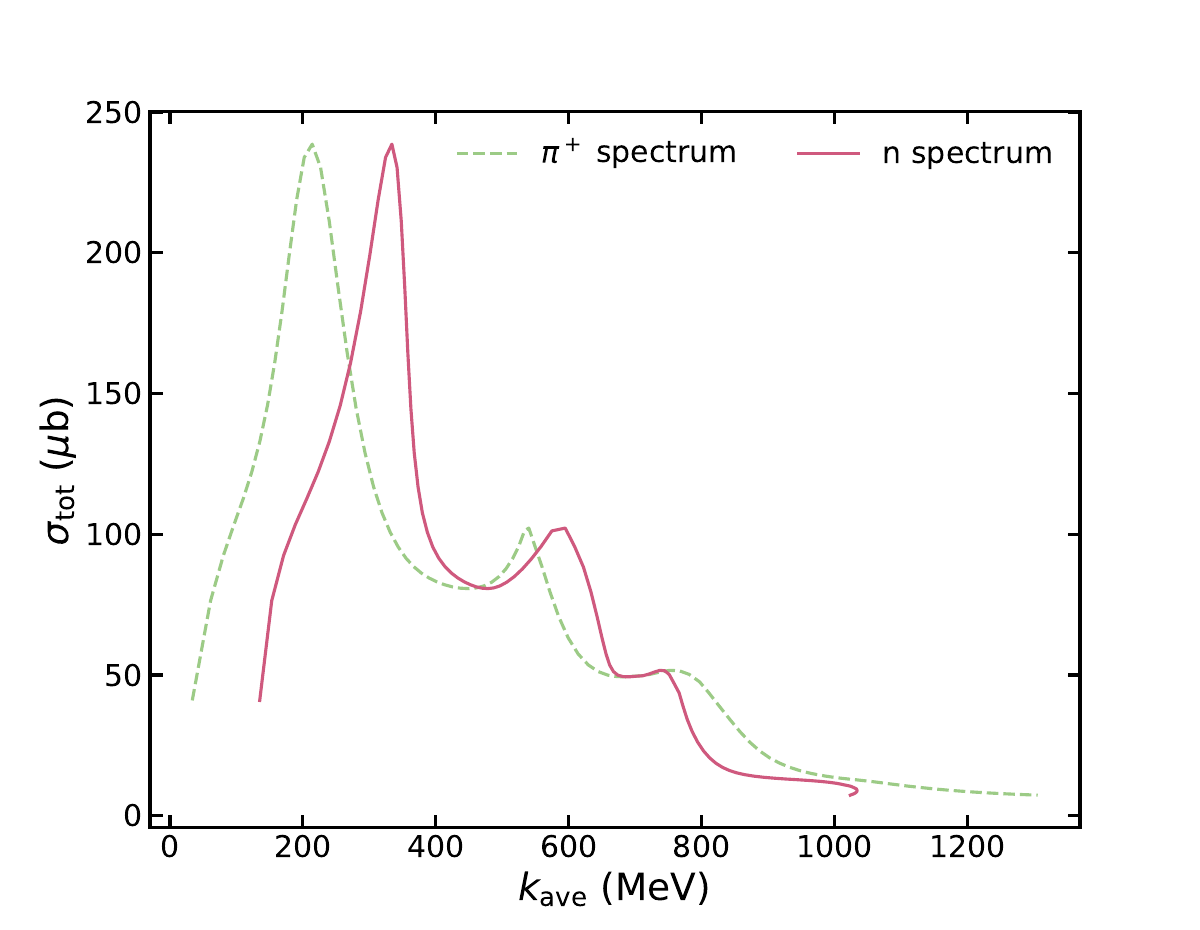}
    \includegraphics[width=0.45\textwidth]{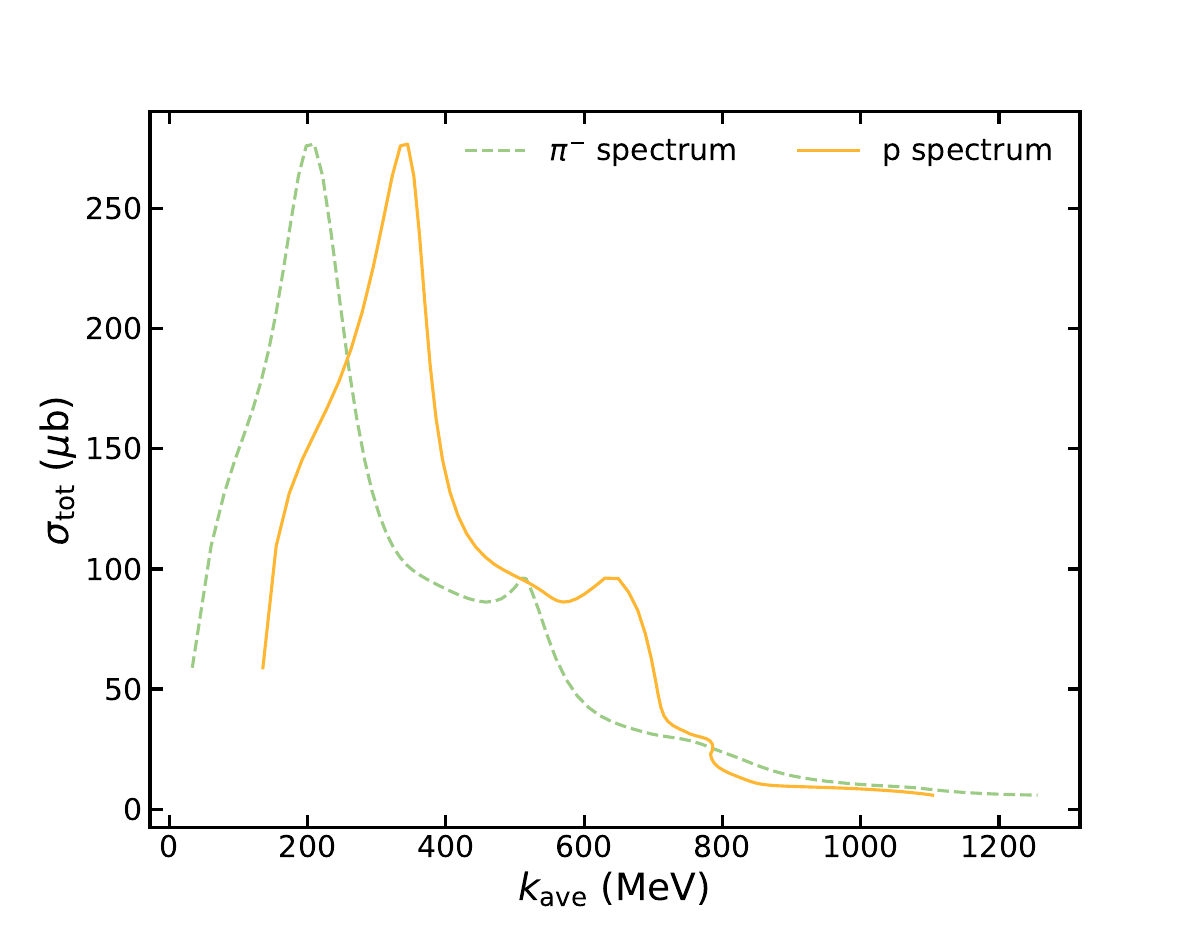}
    \vspace{0.1cm} 
    \includegraphics[width=0.45\textwidth]{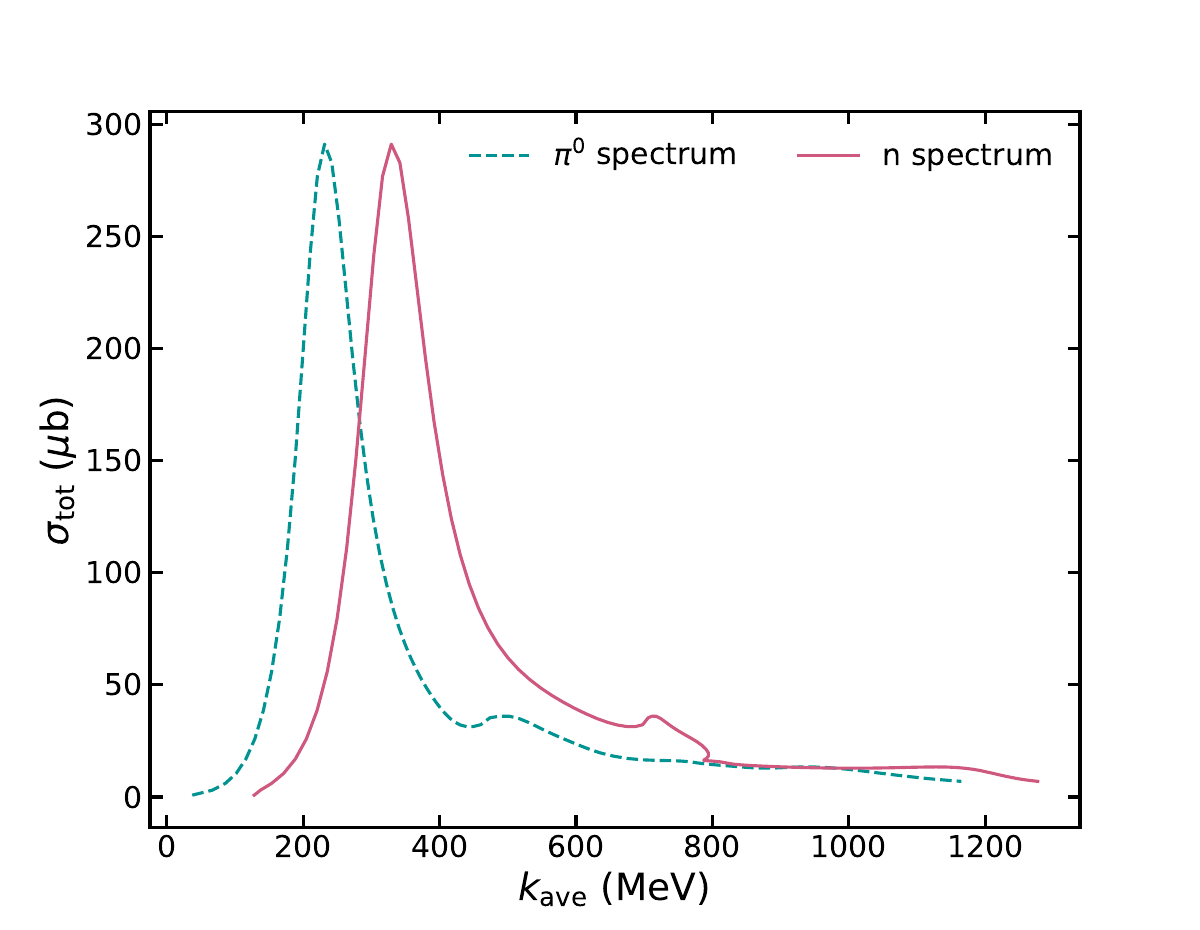}
    \includegraphics[width=0.45\textwidth]{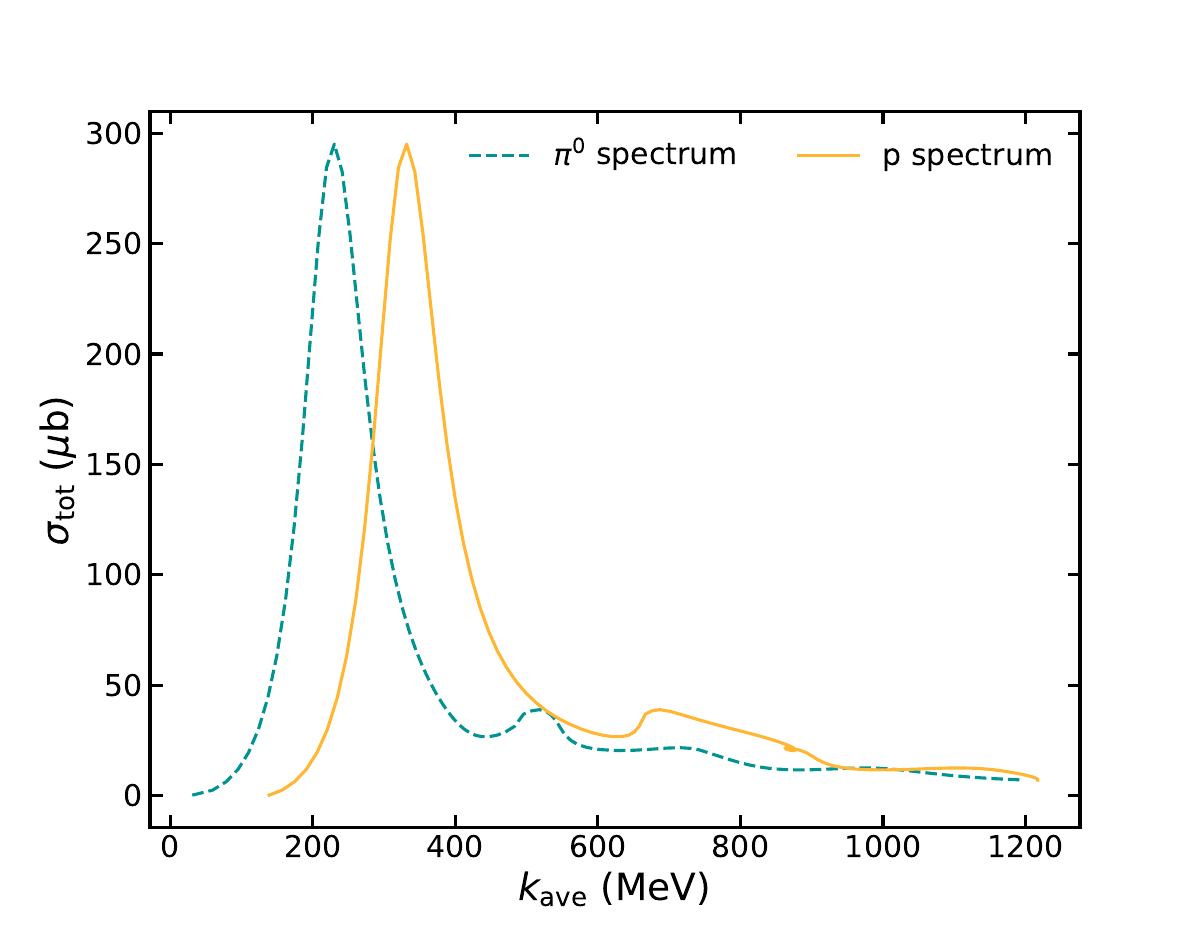}
    
    \caption{(upper left) Averaged spectra associated with the $\gamma + p \rightarrow \pi^{+} + n$ cross section, process (\ref{p:gamma_1m_ncreate}); 
    (upper right) spectra associated with the $\gamma + n \rightarrow \pi^{-} + p$ cross section, process (\ref{p:gamma_1m_nkill}); 
    (bottom left) spectra associated with the $\gamma + n \rightarrow \pi^{0} + n$ cross section, process (\ref{p:gamma_1m_nscatter}); 
    (bottom right) spectra associated with the $\gamma + p \rightarrow \pi^{0} + p$ cross section, process (\ref{p:gamma_1m_pscatter}). 
    }
    \label{fig:specs}
\end{figure}


M.~R.~M. thanks Yong Zhao at Argonne National Laboratory for the introduction to H. Lee. 
M.~R.~M. thanks T.~M.~Sprouse, R.~T.~Wollaeger, and O.~Korobkin for valuable discussions. 
M.~R.~M. acknowledges support from the Directed Asymmetric Network Graphs for Research (DANGR) initiative at Los Alamos National Laboratory (LANL). 
LANL is operated by Triad National Security, LLC, for the National Nuclear Security Administration of U.S. Department of Energy (Contract No. 89233218CNA000001). 
T.-S. H. Lee is supported by the Office of Science, Office of Nuclear Physics, contract no. DE-AC02-06CH11357.
N.L-R. acknowledges support from the Laboratory Directed Research and Development program of LANL project numbers 20230115ER and 20230217ER.
B.~L.~B. acknowledges support from the Advanced Simulation and Computing Program of LANL as a Metropolis Postdoc Fellow.


\bibliographystyle{aasjournal}
\bibliography{refs}

\begin{thebibliography}{}
\expandafter\ifx\csname natexlab\endcsname\relax\def\natexlab#1{#1}\fi
\providecommand{\url}[1]{\href{#1}{#1}}
\providecommand{\dodoi}[1]{doi:~\href{http://doi.org/#1}{\nolinkurl{#1}}}
\providecommand{\doeprint}[1]{\href{http://ascl.net/#1}{\nolinkurl{http://ascl.net/#1}}}
\providecommand{\doarXiv}[1]{\href{https://arxiv.org/abs/#1}{\nolinkurl{https://arxiv.org/abs/#1}}}

\bibitem[{{Abbott} {et~al.}(2017){Abbott}, {Abbott}, {Abbott}, {Acernese},
  {Ackley}, {Adams}, {Adams}, {Addesso}, {Adhikari}, {Adya}, \& et~al.}]{Ab17}
{Abbott}, B.~P., {Abbott}, R., {Abbott}, T.~D., {et~al.} 2017, Physical Review
  Letters, 119, 161101, \dodoi{10.1103/PhysRevLett.119.161101}

\bibitem[{{Ahlgren} {et~al.}(2019){Ahlgren}, {Larsson}, {Valan}, {Mortlock},
  {Ryde}, \& {Pe'er}}]{Ahl19}
{Ahlgren}, B., {Larsson}, J., {Valan}, V., {et~al.} 2019, \apj, 880, 76,
  \dodoi{10.3847/1538-4357/ab271b}

\bibitem[{{Anand} {et~al.}(2024){Anand}, {Barnes}, {Yang}, {Kasliwal},
  {Coughlin}, {Sollerman}, {De}, {Fremling}, {Corsi}, {Ho}, {Balasubramanian},
  {Omand}, {Srinivasaragavan}, {Cenko}, {Ahumada}, {Andreoni}, {Dahiwale},
  {Das}, {Jencson}, {Karambelkar}, {Kumar}, {Metzger}, {Perley}, {Sarin},
  {Schweyer}, {Schulze}, {Sharma}, {Sit}, {Stein}, {Tartaglia}, {Tinyanont},
  {Tzanidakis}, {van Roestel}, {Yao}, {Bloom}, {Cook}, {Dekany}, {Graham},
  {Groom}, {Kaplan}, {Masci}, {Medford}, {Riddle}, \& {Zhang}}]{Anand24}
{Anand}, S., {Barnes}, J., {Yang}, S., {et~al.} 2024, \apj, 962, 68,
  \dodoi{10.3847/1538-4357/ad11df}

\bibitem[{Asano \& Mészáros(2013)}]{Asano2013}
Asano, K., \& Mészáros, P. 2013, Journal of Cosmology and Astroparticle
  Physics, 2013, 008–008, \dodoi{10.1088/1475-7516/2013/09/008}

\bibitem[{Band {et~al.}(1993)Band, Matteson, Ford, Schaefer, Palmer, Teegarden,
  Cline, Briggs, Paciesas, Pendleton, Fishman, Kouveliotou, Meegan, Wilson, \&
  Lestrade}]{Band1993}
Band, D., Matteson, J., Ford, L., {et~al.} 1993, The Astrophysical Journal,
  413, 281, \dodoi{10.1086/172995}

\bibitem[{{Barnes} {et~al.}(2021){Barnes}, {Zhu}, {Lund}, {Sprouse}, {Vassh},
  {McLaughlin}, {Mumpower}, \& {Surman}}]{Barnes2021}
{Barnes}, J., {Zhu}, Y.~L., {Lund}, K.~A., {et~al.} 2021, \apj, 918, 44,
  \dodoi{10.3847/1538-4357/ac0aec}

\bibitem[{Battino {et~al.}(2019)Battino, Tattersall, Lederer-Woods, Herwig,
  Denissenkov, Hirschi, Trappitsch, den Hartogh, Pignatari, \&
  Collaboration†)}]{Battino2019}
Battino, U., Tattersall, A., Lederer-Woods, C., {et~al.} 2019, Monthly Notices
  of the Royal Astronomical Society, 489, 1082, \dodoi{10.1093/mnras/stz2158}

\bibitem[{{Beloborodov}(2010)}]{Bel10}
{Beloborodov}, A.~M. 2010, \mnras, 407, 1033,
  \dodoi{10.1111/j.1365-2966.2010.16770.x}

\bibitem[{{Berezinskii} \& {Smirnov}(1975)}]{Berezinskii1975}
{Berezinskii}, V.~S., \& {Smirnov}, A.~I. 1975, \apss, 32, 461,
  \dodoi{10.1007/BF00643157}

\bibitem[{Bethe \& Brown(1995)}]{Bethe1995}
Bethe, H.~A., \& Brown, G.~E. 1995, Reviews of Modern Physics, 67, 545,
  \dodoi{10.1103/RevModPhys.67.545}

\bibitem[{{Blanchard} {et~al.}(2024){Blanchard}, {Villar}, {Chornock},
  {Laskar}, {Li}, {Leja}, {Pierel}, {Berger}, {Margutti}, {Alexander},
  {Barnes}, {Cendes}, {Eftekhari}, {Kasen}, {LeBaron}, {Metzger}, {Muzerolle
  Page}, {Rest}, {Sears}, {Siegel}, \& {Yadavalli}}]{Blanch24}
{Blanchard}, P.~K., {Villar}, V.~A., {Chornock}, R., {et~al.} 2024, Nature
  Astronomy, 8, 774, \dodoi{10.1038/s41550-024-02237-4}

\bibitem[{{Blandford} \& {Eichler}(1987)}]{Blandford1987}
{Blandford}, R., \& {Eichler}, D. 1987, \physrep, 154, 1,
  \dodoi{10.1016/0370-1573(87)90134-7}

\bibitem[{{Blandford} {et~al.}(2019){Blandford}, {Meier}, \&
  {Readhead}}]{Bland19}
{Blandford}, R., {Meier}, D., \& {Readhead}, A. 2019, \araa, 57, 467,
  \dodoi{10.1146/annurev-astro-081817-051948}

\bibitem[{Blandford \& McKee(1976)}]{Blandford1976}
Blandford, R.~D., \& McKee, C.~F. 1976, The Physics of Fluids, 19, 1130,
  \dodoi{10.1063/1.861619}

\bibitem[{{Blandford} \& {Znajek}(1977)}]{BZ77}
{Blandford}, R.~D., \& {Znajek}, R.~L. 1977, \mnras, 179, 433,
  \dodoi{10.1093/mnras/179.3.433}

\bibitem[{Bohm(1949)}]{Bohm1949}
Bohm, D. 1949, The Characteristics of Electrical Discharges in Magnetic Fields,
  ed. A.~Guthrie (McGraw-Hill)

\bibitem[{{Bromberg} {et~al.}(2011){Bromberg}, {Mikolitzky}, \&
  {Levinson}}]{Brom11a}
{Bromberg}, O., {Mikolitzky}, Z., \& {Levinson}, A. 2011, \apj, 733, 85,
  \dodoi{10.1088/0004-637X/733/2/85}

\bibitem[{Bromberg \& Tchekhovskoy(2015)}]{Bromberg2015}
Bromberg, O., \& Tchekhovskoy, A. 2015, Monthly Notices of the Royal
  Astronomical Society, 456, 1739, \dodoi{10.1093/mnras/stv2591}

\bibitem[{Burbidge {et~al.}(1957)Burbidge, Burbidge, Fowler, \&
  Hoyle}]{Burbidge1957}
Burbidge, E.~M., Burbidge, G.~R., Fowler, W.~A., \& Hoyle, F. 1957, Rev. Mod.
  Phys., 29, 547, \dodoi{10.1103/RevModPhys.29.547}

\bibitem[{{Chi-Kit Cheong} {et~al.}(2024){Chi-Kit Cheong}, {Pitik}, {Longo
  Micchi}, \& {Radice}}]{Cheong24}
{Chi-Kit Cheong}, P., {Pitik}, T., {Longo Micchi}, L.~F., \& {Radice}, D. 2024,
  arXiv e-prints, arXiv:2410.10938, \dodoi{10.48550/arXiv.2410.10938}

\bibitem[{{Choplin} {et~al.}(2021){Choplin}, {Siess}, \&
  {Goriely}}]{Choplin2021}
{Choplin}, A., {Siess}, L., \& {Goriely}, S. 2021, \aap, 648, A119,
  \dodoi{10.1051/0004-6361/202040170}

\bibitem[{{Colgate} \& {Petschek}(1980)}]{Colgate1980}
{Colgate}, S.~A., \& {Petschek}, A.~G. 1980, in Texas Workshop on Type I
  Supernovae, ed. J.~C. {Wheeler}, 42--52

\bibitem[{{C{\^o}t{\'e}} {et~al.}(2018){C{\^o}t{\'e}}, {Denissenkov}, {Herwig},
  {Ruiter}, {Ritter}, {Pignatari}, \& {Belczynski}}]{Cote2018}
{C{\^o}t{\'e}}, B., {Denissenkov}, P., {Herwig}, F., {et~al.} 2018, \apj, 854,
  105, \dodoi{10.3847/1538-4357/aaaae8}

\bibitem[{Cowan {et~al.}(1999)Cowan, Pfeiffer, Kratz, Thielemann, Sneden,
  Burles, Tytler, \& Beers}]{Cowan1999}
Cowan, J.~J., Pfeiffer, B., Kratz, K., {et~al.} 1999, The Astrophysical
  Journal, 521, 194–205, \dodoi{10.1086/307512}

\bibitem[{{Curtis} {et~al.}(2023){Curtis}, {Miller}, {Fr{\"o}hlich}, {Sprouse},
  {Lloyd-Ronning}, \& {Mumpower}}]{Curt23}
{Curtis}, S., {Miller}, J.~M., {Fr{\"o}hlich}, C., {et~al.} 2023, \apjl, 945,
  L13, \dodoi{10.3847/2041-8213/acba16}

\bibitem[{De~Colle {et~al.}(2022)De~Colle, Kumar, \& Hoeflich}]{DeColle2022}
De~Colle, F., Kumar, P., \& Hoeflich, P. 2022, Monthly Notices of the Royal
  Astronomical Society, 512, 3627, \dodoi{10.1093/mnras/stac742}

\bibitem[{{De Colle} {et~al.}(2018){De Colle}, {Lu}, {Kumar}, {Ramirez-Ruiz},
  \& {Smoot}}]{DeColle18}
{De Colle}, F., {Lu}, W., {Kumar}, P., {Ramirez-Ruiz}, E., \& {Smoot}, G. 2018,
  \mnras, 478, 4553, \dodoi{10.1093/mnras/sty1282}

\bibitem[{Denton \& Tamborra(2018)}]{Denton2018}
Denton, P.~B., \& Tamborra, I. 2018, The Astrophysical Journal, 855, 37,
  \dodoi{10.3847/1538-4357/aaab4a}

\bibitem[{{Duncan} \& {Thompson}(1992)}]{DT92}
{Duncan}, R.~C., \& {Thompson}, C. 1992, \apjl, 392, L9, \dodoi{10.1086/186413}

\bibitem[{{Farouqi, K.} {et~al.}(2022){Farouqi, K.}, {Thielemann, F.-K.},
  {Rosswog, S.}, \& {Kratz, K.-L.}}]{Farouqi2022}
{Farouqi, K.}, {Thielemann, F.-K.}, {Rosswog, S.}, \& {Kratz, K.-L.} 2022,
  A\&A, 663, A70, \dodoi{10.1051/0004-6361/202141038}

\bibitem[{{Fender} {et~al.}(2004){Fender}, {Belloni}, \& {Gallo}}]{Fend04}
{Fender}, R.~P., {Belloni}, T.~M., \& {Gallo}, E. 2004, \mnras, 355, 1105,
  \dodoi{10.1111/j.1365-2966.2004.08384.x}

\bibitem[{{Freiburghaus} {et~al.}(1999){Freiburghaus}, {Rosswog}, \&
  {Thielemann}}]{Freiburghaus1999}
{Freiburghaus}, C., {Rosswog}, S., \& {Thielemann}, F.~K. 1999, \apjl, 525,
  L121, \dodoi{10.1086/312343}

\bibitem[{{Fujimoto} {et~al.}(2007){Fujimoto}, {Hashimoto}, {Kotake}, \&
  {Yamada}}]{Fujimoto2007}
{Fujimoto}, S.-i., {Hashimoto}, M.-a., {Kotake}, K., \& {Yamada}, S. 2007,
  \apj, 656, 382, \dodoi{10.1086/509908}

\bibitem[{{Fuller} \& {Ro}(2018)}]{fuller18}
{Fuller}, J., \& {Ro}, S. 2018, \mnras, 476, 1853, \dodoi{10.1093/mnras/sty369}

\bibitem[{{Gehrels} {et~al.}(2009){Gehrels}, {Ramirez-Ruiz}, \& {Fox}}]{GRRF09}
{Gehrels}, N., {Ramirez-Ruiz}, E., \& {Fox}, D.~B. 2009, \araa, 47, 567,
  \dodoi{10.1146/annurev.astro.46.060407.145147}

\bibitem[{{Giannios} \& {Spruit}(2007{\natexlab{a}})}]{Giannios2007}
{Giannios}, D., \& {Spruit}, H.~C. 2007{\natexlab{a}}, \aap, 469, 1,
  \dodoi{10.1051/0004-6361:20066739}

\bibitem[{{Giannios} \& {Spruit}(2007{\natexlab{b}})}]{GS07}
---. 2007{\natexlab{b}}, \aap, 469, 1, \dodoi{10.1051/0004-6361:20066739}

\bibitem[{{Ginzburg} \& {Syrovatskii}(1964)}]{Ginzburg1964}
{Ginzburg}, V.~L., \& {Syrovatskii}, S.~I. 1964, {The Origin of Cosmic Rays}

\bibitem[{{Gottlieb} {et~al.}(2022){Gottlieb}, {Lalakos}, {Bromberg}, {Liska},
  \& {Tchekhovskoy}}]{Gott22}
{Gottlieb}, O., {Lalakos}, A., {Bromberg}, O., {Liska}, M., \& {Tchekhovskoy},
  A. 2022, \mnras, 510, 4962, \dodoi{10.1093/mnras/stab3784}

\bibitem[{Gottlieb {et~al.}(2020)Gottlieb, Nakar, \& Bromberg}]{Gottlieb2020}
Gottlieb, O., Nakar, E., \& Bromberg, O. 2020, Monthly Notices of the Royal
  Astronomical Society, 500, 3511–3526, \dodoi{10.1093/mnras/staa3501}

\bibitem[{Guetta {et~al.}(2004)Guetta, Hooper, Alvarez-Muñiz, Halzen, \&
  Reuveni}]{Guetta2004}
Guetta, D., Hooper, D., Alvarez-Muñiz, J., Halzen, F., \& Reuveni, E. 2004,
  Astroparticle Physics, 20, 429–455, \dodoi{10.1016/s0927-6505(03)00211-1}

\bibitem[{Halevi {et~al.}(2023)Halevi, Wu, Mösta, Gottlieb, Tchekhovskoy, \&
  Aguilera-Dena}]{Halevi2023}
Halevi, G., Wu, B., Mösta, P., {et~al.} 2023, The Astrophysical Journal
  Letters, 944, L38, \dodoi{10.3847/2041-8213/acb702}

\bibitem[{{Harrison} \& {Kobayashi}(2013)}]{HK13}
{Harrison}, R., \& {Kobayashi}, S. 2013, \apj, 772, 101,
  \dodoi{10.1088/0004-637X/772/2/101}

\bibitem[{{Herwig} {et~al.}(2014){Herwig}, {Woodward}, {Lin}, {Knox}, \&
  {Fryer}}]{herwig14}
{Herwig}, F., {Woodward}, P.~R., {Lin}, P.-H., {Knox}, M., \& {Fryer}, C. 2014,
  \apjl, 792, L3, \dodoi{10.1088/2041-8205/792/1/L3}

\bibitem[{Herwig {et~al.}(2008)Herwig, Bennett, Diehl, Fryer, Hirschi,
  Hungerford, Magkotsios, Pignatari, Rockefeller, Timmes, \&
  Young}]{Herwig2008}
Herwig, F., Bennett, M., Diehl, S., {et~al.} 2008, Nucleosynthesis simulations
  for a wide range of nuclear production sites from NuGrid.
\newblock \doarXiv{0811.4653}

\bibitem[{{Hillas}(1984)}]{Hillas1984}
{Hillas}, A.~M. 1984, \araa, 22, 425,
  \dodoi{10.1146/annurev.aa.22.090184.002233}

\bibitem[{{Hotokezaka} {et~al.}(2018){Hotokezaka}, {Beniamini}, \&
  {Piran}}]{Hoto18}
{Hotokezaka}, K., {Beniamini}, P., \& {Piran}, T. 2018, International Journal
  of Modern Physics D, 27, 1842005, \dodoi{10.1142/S0218271818420051}

\bibitem[{Hümmer {et~al.}(2012)Hümmer, Baerwald, \& Winter}]{Hummer2012}
Hümmer, S., Baerwald, P., \& Winter, W. 2012, Physical Review Letters, 108,
  \dodoi{10.1103/physrevlett.108.231101}

\bibitem[{{Izzo} {et~al.}(2019){Izzo}, {de Ugarte Postigo}, {Maeda},
  {Th{\"o}ne}, {Kann}, {Della Valle}, {Sagues Carracedo}, {Micha{\l}owski},
  {Schady}, {Schmidl}, {Selsing}, {Starling}, {Suzuki}, {Bensch}, {Bolmer},
  {Campana}, {Cano}, {Covino}, {Fynbo}, {Hartmann}, {Heintz}, {Hjorth},
  {Japelj}, {Kami{\'n}ski}, {Kaper}, {Kouveliotou}, {Kru{\.Z}y{\'n}ski},
  {Kwiatkowski}, {Leloudas}, {Levan}, {Malesani}, {Micha{\l}owski},
  {Piranomonte}, {Pugliese}, {Rossi}, {S{\'a}nchez-Ram{\'\i}rez}, {Schulze},
  {Steeghs}, {Tanvir}, {Ulaczyk}, {Vergani}, \& {Wiersema}}]{Izzo19}
{Izzo}, L., {de Ugarte Postigo}, A., {Maeda}, K., {et~al.} 2019, \nat, 565,
  324, \dodoi{10.1038/s41586-018-0826-3}

\bibitem[{{Jiang}(2023)}]{Jiang23}
{Jiang}, Y.-F. 2023, Galaxies, 11, 105, \dodoi{10.3390/galaxies11050105}

\bibitem[{Juliá-Díaz {et~al.}(2007)Juliá-Díaz, Lee, Matsuyama, \&
  Sato}]{JuliaDiaz2007}
Juliá-Díaz, B., Lee, T.-S.~H., Matsuyama, A., \& Sato, T. 2007, Physical
  Review C, 76, \dodoi{10.1103/physrevc.76.065201}

\bibitem[{{Just} {et~al.}(2022){Just}, {Aloy}, {Obergaulinger}, \&
  {Nagataki}}]{Just22}
{Just}, O., {Aloy}, M.~A., {Obergaulinger}, M., \& {Nagataki}, S. 2022, \apjl,
  934, L30, \dodoi{10.3847/2041-8213/ac83a1}

\bibitem[{Kamano(2017)}]{Kamano2017b}
Kamano, H. 2017, in Proceedings of the 14th International Conference on
  Meson-Nucleon Physics and the Structure of the Nucleon (MENU2016) (Journal of
  the Physical Society of Japan), \dodoi{10.7566/jpscp.13.010012}

\bibitem[{Kamano {et~al.}(2019)Kamano, Lee, Nakamura, \& Sato}]{Kamano2019}
Kamano, H., Lee, T. S.~H., Nakamura, S.~X., \& Sato, T. 2019, The ANL-Osaka
  Partial-Wave Amplitudes of $\pi N$ and $\gamma N$ Reactions.
\newblock \doarXiv{1909.11935}

\bibitem[{Kamano {et~al.}(2010)Kamano, Nakamura, Lee, \& Sato}]{Kamano2010}
Kamano, H., Nakamura, S.~X., Lee, T.-S.~H., \& Sato, T. 2010, Phys. Rev. C, 81,
  065207, \dodoi{10.1103/PhysRevC.81.065207}

\bibitem[{Kamano {et~al.}(2013)Kamano, Nakamura, Lee, \& Sato}]{Kamano2013}
---. 2013, Phys. Rev. C, 88, 035209, \dodoi{10.1103/PhysRevC.88.035209}

\bibitem[{{Kargaltsev} {et~al.}(2015){Kargaltsev}, {Cerutti}, {Lyubarsky}, \&
  {Striani}}]{Kar15}
{Kargaltsev}, O., {Cerutti}, B., {Lyubarsky}, Y., \& {Striani}, E. 2015, \ssr,
  191, 391, \dodoi{10.1007/s11214-015-0171-x}

\bibitem[{Kippenhahn {et~al.}(2012)Kippenhahn, Weigert, \&
  Weiss}]{Kippenhahn2012}
Kippenhahn, R., Weigert, A., \& Weiss, A. 2012, Stellar Structure and
  Evolution, 2nd edn. (Berlin Heidelberg: Springer-Verlag)

\bibitem[{{K{\"o}lligan} \& {Kuiper}(2018)}]{Koll18}
{K{\"o}lligan}, A., \& {Kuiper}, R. 2018, \aap, 620, A182,
  \dodoi{10.1051/0004-6361/201833686}

\bibitem[{{Korobkin} {et~al.}(2020){Korobkin}, {Hungerford}, {Fryer},
  {Mumpower}, {Misch}, {Sprouse}, {Lippuner}, {Surman}, {Couture}, {Bloser},
  {Shirazi}, {Even}, {Vestrand}, \& {Miller}}]{Korobkin2020}
{Korobkin}, O., {Hungerford}, A.~M., {Fryer}, C.~L., {et~al.} 2020, \apj, 889,
  168, \dodoi{10.3847/1538-4357/ab64d8}

\bibitem[{{Kumar} \& {Zhang}(2015)}]{KZ15}
{Kumar}, P., \& {Zhang}, B. 2015, \physrep, 561, 1,
  \dodoi{10.1016/j.physrep.2014.09.008}

\bibitem[{{Lattimer} \& {Schramm}(1974)}]{Lattimer1974}
{Lattimer}, J.~M., \& {Schramm}, D.~N. 1974, \apjl, 192, L145,
  \dodoi{10.1086/181612}

\bibitem[{{Lazzati} {et~al.}(2015){Lazzati}, {Morsony}, \&
  {L{\'o}pez-C{\'a}mara}}]{Laz15}
{Lazzati}, D., {Morsony}, B.~J., \& {L{\'o}pez-C{\'a}mara}, D. 2015, Journal of
  High Energy Astrophysics, 7, 17, \dodoi{10.1016/j.jheap.2015.04.001}

\bibitem[{{Lazzati} {et~al.}(2023){Lazzati}, {Perna}, {Gompertz}, \&
  {Levan}}]{Laz23}
{Lazzati}, D., {Perna}, R., {Gompertz}, B.~P., \& {Levan}, A.~J. 2023, \apjl,
  950, L20, \dodoi{10.3847/2041-8213/acd18c}

\bibitem[{Lee(2019)}]{Lee2019}
Lee, T.-S.~H. 2019.
\newblock \url{https://www.phy.anl.gov/theory/research/anl-osaka-pwa}

\bibitem[{{Levan} {et~al.}(2016){Levan}, {Crowther}, {de Grijs}, {Langer},
  {Xu}, \& {Yoon}}]{Lev16}
{Levan}, A., {Crowther}, P., {de Grijs}, R., {et~al.} 2016, \ssr, 202, 33,
  \dodoi{10.1007/s11214-016-0312-x}

\bibitem[{{Lewin} \& {van der Klis}(2010)}]{Lew10}
{Lewin}, W., \& {van der Klis}, M. 2010, {Compact Stellar X-ray Sources}

\bibitem[{Lippuner \& Roberts(2017)}]{Lippuner2017}
Lippuner, J., \& Roberts, L.~F. 2017, The Astrophysical Journal Supplement
  Series, 233, 18, \dodoi{10.3847/1538-4365/aa94cb}

\bibitem[{{Lithwick} \& {Sari}(2001)}]{LS01}
{Lithwick}, Y., \& {Sari}, R. 2001, \apj, 555, 540, \dodoi{10.1086/321455}

\bibitem[{{Lloyd-Ronning} {et~al.}(2019){Lloyd-Ronning}, {Fryer}, {Miller},
  {Prasad}, {Torres}, \& {Martin}}]{LR19}
{Lloyd-Ronning}, N.~M., {Fryer}, C., {Miller}, J.~M., {et~al.} 2019, \mnras,
  485, 203, \dodoi{10.1093/mnras/stz390}

\bibitem[{Longo \& Moyer(1962)}]{Longo1962}
Longo, M.~J., \& Moyer, B.~J. 1962, Phys. Rev., 125, 701,
  \dodoi{10.1103/PhysRev.125.701}

\bibitem[{{MacDonald} \& {Thorne}(1982)}]{MT82}
{MacDonald}, D., \& {Thorne}, K.~S. 1982, \mnras, 198, 345,
  \dodoi{10.1093/mnras/198.2.345}

\bibitem[{{MacFadyen} \& {Woosley}(1999)}]{MW98}
{MacFadyen}, A.~I., \& {Woosley}, S.~E. 1999, \apj, 524, 262,
  \dodoi{10.1086/307790}

\bibitem[{Mannheim(1995)}]{Mannheim1995}
Mannheim, K. 1995, Astroparticle Physics, 3, 295,
  \dodoi{https://doi.org/10.1016/0927-6505(94)00044-4}

\bibitem[{Maria(2016)}]{Maria2016}
Maria, L. 2016, Journal of Physics: Conference Series, 703, 012003,
  \dodoi{10.1088/1742-6596/703/1/012003}

\bibitem[{Mastichiadis \& Petropoulou(2021)}]{Mastichiadis2021}
Mastichiadis, A., \& Petropoulou, M. 2021, The Astrophysical Journal, 906, 131,
  \dodoi{10.3847/1538-4357/abc952}

\bibitem[{Matsuyama {et~al.}(2007)Matsuyama, Sato, \& Lee}]{Matsuyama2007}
Matsuyama, A., Sato, T., \& Lee, T.-S. 2007, Physics Reports, 439, 193–253,
  \dodoi{10.1016/j.physrep.2006.12.003}

\bibitem[{{Mauerhan} {et~al.}(2018){Mauerhan}, {Filippenko}, {Zheng}, {Brink},
  {Graham}, {Shivvers}, \& {Clubb}}]{mauerhan18}
{Mauerhan}, J.~C., {Filippenko}, A.~V., {Zheng}, W., {et~al.} 2018, \mnras,
  478, 5050, \dodoi{10.1093/mnras/sty1307}

\bibitem[{{Mesler} {et~al.}(2012){Mesler}, {Whalen}, {Lloyd-Ronning}, {Fryer},
  \& {Pihlstr{\"o}m}}]{Mes12}
{Mesler}, R.~A., {Whalen}, D.~J., {Lloyd-Ronning}, N.~M., {Fryer}, C.~L., \&
  {Pihlstr{\"o}m}, Y.~M. 2012, \apj, 757, 117,
  \dodoi{10.1088/0004-637X/757/2/117}

\bibitem[{{M{\'e}sz{\'a}ros}(2006)}]{Mesz06}
{M{\'e}sz{\'a}ros}, P. 2006, Reports on Progress in Physics, 69, 2259,
  \dodoi{10.1088/0034-4885/69/8/R01}

\bibitem[{Metzger(2019)}]{Metzger2019}
Metzger, B.~D. 2019, Living Reviews in Relativity, 23,
  \dodoi{10.1007/s41114-019-0024-0}

\bibitem[{Metzger {et~al.}(2014)Metzger, Bauswein, Goriely, \&
  Kasen}]{Metzger2014}
Metzger, B.~D., Bauswein, A., Goriely, S., \& Kasen, D. 2014, Monthly Notices
  of the Royal Astronomical Society, 446, 1115–1120,
  \dodoi{10.1093/mnras/stu2225}

\bibitem[{{Meyer} {et~al.}(1998){Meyer}, {Krishnan}, \& {Clayton}}]{Meyer1998}
{Meyer}, B.~S., {Krishnan}, T.~D., \& {Clayton}, D.~D. 1998, \apj, 498, 808,
  \dodoi{10.1086/305562}

\bibitem[{{Miller} {et~al.}(2020){Miller}, {Sprouse}, {Fryer}, {Ryan},
  {Dolence}, {Mumpower}, \& {Surman}}]{Miller2020}
{Miller}, J.~M., {Sprouse}, T.~M., {Fryer}, C.~L., {et~al.} 2020, \apj, 902,
  66, \dodoi{10.3847/1538-4357/abb4e3}

\bibitem[{{Moriya} {et~al.}(2017){Moriya}, {Yoon}, {Gr{\"a}fener}, \&
  {Blinnikov}}]{Moriya2017}
{Moriya}, T.~J., {Yoon}, S.-C., {Gr{\"a}fener}, G., \& {Blinnikov}, S.~I. 2017,
  \mnras, 469, L108, \dodoi{10.1093/mnrasl/slx056}

\bibitem[{{Morozova} {et~al.}(2018){Morozova}, {Piro}, \&
  {Valenti}}]{Morozova2018}
{Morozova}, V., {Piro}, A.~L., \& {Valenti}, S. 2018, \apj, 858, 15,
  \dodoi{10.3847/1538-4357/aab9a6}

\bibitem[{{Morsony} {et~al.}(2010){Morsony}, {Lazzati}, \& {Begelman}}]{MLB10}
{Morsony}, B.~J., {Lazzati}, D., \& {Begelman}, M.~C. 2010, \apj, 723, 267,
  \dodoi{10.1088/0004-637X/723/1/267}

\bibitem[{{M{\"u}cke} {et~al.}(1999){M{\"u}cke}, {Rachen}, {Engel},
  {Protheroe}, \& {Stanev}}]{Mucke1999}
{M{\"u}cke}, A., {Rachen}, J.~P., {Engel}, R., {Protheroe}, R.~J., \& {Stanev},
  T. 1999, \pasa, 16, 160, \dodoi{10.1071/AS99160}

\bibitem[{Mumpower {et~al.}(2016)Mumpower, Surman, McLaughlin, \&
  Aprahamian}]{Mumpower2016}
Mumpower, M., Surman, R., McLaughlin, G., \& Aprahamian, A. 2016, Progress in
  Particle and Nuclear Physics, 86, 86–126,
  \dodoi{10.1016/j.ppnp.2015.09.001}

\bibitem[{{Mumpower} {et~al.}(2024){Mumpower}, {Sprouse}, {Miller}, {Lund},
  {Garcia}, {Vassh}, {McLaughlin}, \& {Surman}}]{Mumpower2024}
{Mumpower}, M.~R., {Sprouse}, T.~M., {Miller}, J.~M., {et~al.} 2024, \apj, 970,
  173, \dodoi{10.3847/1538-4357/ad5afc}

\bibitem[{Murase \& Ioka(2013)}]{Murase2013b}
Murase, K., \& Ioka, K. 2013, Physical Review Letters, 111,
  \dodoi{10.1103/physrevlett.111.121102}

\bibitem[{Murase {et~al.}(2013)Murase, Kashiyama, \& Mészáros}]{Murase2013}
Murase, K., Kashiyama, K., \& Mészáros, P. 2013, Physical Review Letters,
  111, \dodoi{10.1103/physrevlett.111.131102}

\bibitem[{Mészáros \& Waxman(2001)}]{Meszaros2001}
Mészáros, P., \& Waxman, E. 2001, Physical Review Letters, 87,
  \dodoi{10.1103/physrevlett.87.171102}

\bibitem[{Nakamura {et~al.}(2018)Nakamura, Kamano, Lee, \& Sato}]{Nakamura2018}
Nakamura, S.~X., Kamano, H., Lee, T. S.~H., \& Sato, T. 2018, Nuclear
  applications of ANL-Osaka amplitudes: pion photo-productions on deuteron.
\newblock \doarXiv{1804.04757}

\bibitem[{{Nakar} \& {Piran}(2017)}]{NP18}
{Nakar}, E., \& {Piran}, T. 2017, \apj, 834, 28,
  \dodoi{10.3847/1538-4357/834/1/28}

\bibitem[{Oka {et~al.}(2024)Oka, Makishima, \& Terasawa}]{Oka2024}
Oka, M., Makishima, K., \& Terasawa, T. 2024, Maximum Energy of Particles in
  Plasmas.
\newblock \doarXiv{2412.00564}

\bibitem[{Oppenheimer \& Volkoff(1939)}]{Oppenheimer1939}
Oppenheimer, J.~R., \& Volkoff, G.~M. 1939, Physical Review, 55, 374,
  \dodoi{10.1103/PhysRev.55.374}

\bibitem[{{Particle Data Group} {et~al.}(2022){Particle Data Group}, Workman,
  Burkert, Crede, Klempt, Thoma, Tiator, Agashe, Aielli, Allanach, Amsler,
  Antonelli, Aschenauer, Asner, Baer, Banerjee, Barnett, Baudis, Bauer, Beatty,
  Belousov, Beringer, Bettini, Biebel, Black, Blucher, Bonventre, Bryzgalov,
  Buchmuller, Bychkov, Cahn, Carena, Ceccucci, Cerri, Chivukula, Cowan,
  Cranmer, Cremonesi, D'Ambrosio, Damour, de~Florian, de~Gouvêa, DeGrand,
  de~Jong, Demers, Dobrescu, D'Onofrio, Doser, Dreiner, Eerola, Egede,
  Eidelman, El-Khadra, Ellis, Eno, Erler, Ezhela, Fetscher, Fields, Freitas,
  Gallagher, Gershtein, Gherghetta, Gonzalez-Garcia, Goodman, Grab, Gritsan,
  Grojean, Groom, Grünewald, Gurtu, Gutsche, Haber, Hamel, Hanhart, Hashimoto,
  Hayato, Hebecker, Heinemeyer, Hernández-Rey, Hikasa, Hisano, Höcker,
  Holder, Hsu, Huston, Hyodo, Ianni, Kado, Karliner, Katz, Kenzie, Khoze,
  Klein, Krauss, Kreps, Križan, Krusche, Kwon, Lahav, Laiho, Lellouch,
  Lesgourgues, Liddle, Ligeti, Lin, Lippmann, Liss, Littenberg, Lourenço,
  Lugovsky, Lugovsky, Lusiani, Makida, Maltoni, Mannel, Manohar, Marciano,
  Masoni, Matthews, Meißner, Melzer-Pellmann, Mikhasenko, Miller, Milstead,
  Mitchell, Mönig, Molaro, Moortgat, Moskovic, Nakamura, Narain, Nason, Navas,
  Nelles, Neubert, Nevski, Nir, Olive, Patrignani, Peacock, Petrov, Pianori,
  Pich, Piepke, Pietropaolo, Pomarol, Pordes, Profumo, Quadt, Rabbertz,
  Rademacker, Raffelt, Ramsey-Musolf, Ratcliff, Richardson, Ringwald, Robinson,
  Roesler, Rolli, Romaniouk, Rosenberg, Rosner, Rybka, Ryskin, Ryutin, Sakai,
  Sarkar, Sauli, Schneider, Schönert, Scholberg, Schwartz, Schwiening, Scott,
  Sefkow, Seljak, Sharma, Sharpe, Shiltsev, Signorelli, Silari, Simon,
  Sjöstrand, Skands, Skwarnicki, Smoot, Soffer, Sozzi, Spanier, Spiering,
  Stahl, Stone, Sumino, Syphers, Takahashi, Tanabashi, Tanaka, Taševský,
  Terao, Terashi, Terning, Thorne, Titov, Tkachenko, Tovey, Trabelsi, Urquijo,
  Valencia, Van~de Water, Varelas, Venanzoni, Verde, Vivarelli, Vogel,
  Vogelsang, Vorobyev, Wakely, Walkowiak, Walter, Wands, Weinberg, Weinberg,
  Wermes, White, Wiencke, Willocq, Wohl, Woody, Yao, Yokoyama, Yoshida,
  Zanderighi, Zeller, Zenin, Zhu, Zhu, Zimmermann, \& Zyla}]{PDG2022}
{Particle Data Group}, Workman, R.~L., Burkert, V.~D., {et~al.} 2022, Progress
  of Theoretical and Experimental Physics, 2022, 083C01,
  \dodoi{10.1093/ptep/ptac097}

\bibitem[{{Pe'er} {et~al.}(2006){Pe'er}, {M{\'e}sz{\'a}ros}, \&
  {Rees}}]{Peer06}
{Pe'er}, A., {M{\'e}sz{\'a}ros}, P., \& {Rees}, M.~J. 2006, \apj, 642, 995,
  \dodoi{10.1086/501424}

\bibitem[{Piran(1999)}]{Piran1999}
Piran, T. 1999, Physics Reports, 314, 575,
  \dodoi{https://doi.org/10.1016/S0370-1573(98)00127-6}

\bibitem[{{Piran}(2004)}]{Pir04}
{Piran}, T. 2004, Reviews of Modern Physics, 76, 1143,
  \dodoi{10.1103/RevModPhys.76.1143}

\bibitem[{{Piran}(2005)}]{Pir05}
{Piran}, T. 2005, in American Institute of Physics Conference Series, Vol. 784,
  Magnetic Fields in the Universe: From Laboratory and Stars to Primordial
  Structures., ed. E.~M. {de Gouveia dal Pino}, G.~{Lugones}, \& A.~{Lazarian}
  (AIP), 164--174, \dodoi{10.1063/1.2077181}

\bibitem[{{Ramirez-Ruiz} {et~al.}(2002){Ramirez-Ruiz}, {Celotti}, \&
  {Rees}}]{RR02}
{Ramirez-Ruiz}, E., {Celotti}, A., \& {Rees}, M.~J. 2002, \mnras, 337, 1349,
  \dodoi{10.1046/j.1365-8711.2002.05995.x}

\bibitem[{{Rastinejad} {et~al.}(2022){Rastinejad}, {Gompertz}, {Levan}, {Fong},
  {Nicholl}, {Lamb}, {Malesani}, {Nugent}, {Oates}, {Tanvir}, {de Ugarte
  Postigo}, {Kilpatrick}, {Moore}, {Metzger}, {Ravasio}, {Rossi}, {Schroeder},
  {Jencson}, {Sand}, {Smith}, {Ag{\"u}{\'\i} Fern{\'a}ndez}, {Berger},
  {Blanchard}, {Chornock}, {Cobb}, {De Pasquale}, {Fynbo}, {Izzo}, {Kann},
  {Laskar}, {Marini}, {Paterson}, {Escorial}, {Sears}, \&
  {Th{\"o}ne}}]{Rastinejad2022}
{Rastinejad}, J.~C., {Gompertz}, B.~P., {Levan}, A.~J., {et~al.} 2022, \nat,
  612, 223, \dodoi{10.1038/s41586-022-05390-w}

\bibitem[{{Ray} \& {Ferreira}(2021)}]{Ray21}
{Ray}, T.~P., \& {Ferreira}, J. 2021, \nar, 93, 101615,
  \dodoi{10.1016/j.newar.2021.101615}

\bibitem[{Razzaque {et~al.}(2004)Razzaque, Meszaros, \& Zhang}]{Razzaque2004}
Razzaque, S., Meszaros, P., \& Zhang, B. 2004, The Astrophysical Journal, 613,
  1072–1078, \dodoi{10.1086/423166}

\bibitem[{{Roederer} {et~al.}(2023){Roederer}, {Vassh}, {Holmbeck}, {Mumpower},
  {Surman}, {Cowan}, {Beers}, {Ezzeddine}, {Frebel}, {Hansen}, {Placco}, \&
  {Sakari}}]{Roederer2023}
{Roederer}, I.~U., {Vassh}, N., {Holmbeck}, E.~M., {et~al.} 2023, Science, 382,
  1177, \dodoi{10.1126/science.adf1341}

\bibitem[{{Rosswog} {et~al.}(2018){Rosswog}, {Sollerman}, {Feindt}, {Goobar},
  {Korobkin}, {Wollaeger}, {Fremling}, \& {Kasliwal}}]{Rosswog2018}
{Rosswog}, S., {Sollerman}, J., {Feindt}, U., {et~al.} 2018, \aap, 615, A132,
  \dodoi{10.1051/0004-6361/201732117}

\bibitem[{{Sakari} {et~al.}(2018){Sakari}, {Placco}, {Farrell}, {Roederer},
  {Wallerstein}, {Beers}, {Ezzeddine}, {Frebel}, {Hansen}, {Holmbeck},
  {Sneden}, {Cowan}, {Venn}, {Davis}, {Matijevi{\v{c}}}, {Wyse},
  {Bland-Hawthorn}, {Chiappini}, {Freeman}, {Gibson}, {Grebel}, {Helmi},
  {Kordopatis}, {Kunder}, {Navarro}, {Reid}, {Seabroke}, {Steinmetz}, \&
  {Watson}}]{Sakari2018}
{Sakari}, C.~M., {Placco}, V.~M., {Farrell}, E.~M., {et~al.} 2018, \apj, 868,
  110, \dodoi{10.3847/1538-4357/aae9df}

\bibitem[{{Salafia} {et~al.}(2020){Salafia}, {Barbieri}, {Ascenzi}, \&
  {Toffano}}]{Sal20}
{Salafia}, O.~S., {Barbieri}, C., {Ascenzi}, S., \& {Toffano}, M. 2020, \aap,
  636, A105, \dodoi{10.1051/0004-6361/201936335}

\bibitem[{Sato \& Lee(1996)}]{Sato1996}
Sato, T., \& Lee, T.-S.~H. 1996, Phys. Rev. C, 54, 2660,
  \dodoi{10.1103/PhysRevC.54.2660}

\bibitem[{Schatz {et~al.}(2001)Schatz, Aprahamian, Barnard, Bildsten, Cumming,
  Ouellette, Rauscher, Thielemann, \& Wiescher}]{Schatz2001}
Schatz, H., Aprahamian, A., Barnard, V., {et~al.} 2001, Phys. Rev. Lett., 86,
  3471, \dodoi{10.1103/PhysRevLett.86.3471}

\bibitem[{Senno {et~al.}(2016)Senno, Murase, \& Mészáros}]{Senno2016}
Senno, N., Murase, K., \& Mészáros, P. 2016, Physical Review D, 93,
  \dodoi{10.1103/physrevd.93.083003}

\bibitem[{Serebrov \& Fomin(2011)}]{Serebrov2011}
Serebrov, A., \& Fomin, A. 2011, Physics Procedia, 17, 199,
  \dodoi{https://doi.org/10.1016/j.phpro.2011.06.037}

\bibitem[{{Siegel} {et~al.}(2019){Siegel}, {Barnes}, \& {Metzger}}]{Siegel19}
{Siegel}, D.~M., {Barnes}, J., \& {Metzger}, B.~D. 2019, \nat, 569, 241,
  \dodoi{10.1038/s41586-019-1136-0}

\bibitem[{{Smith} \& {Owocki}(2006)}]{smith06}
{Smith}, N., \& {Owocki}, S.~P. 2006, \apjl, 645, L45, \dodoi{10.1086/506523}

\bibitem[{Spitzer(1960)}]{Spitzer1960}
Spitzer, Lyman, J. 1960, The Physics of Fluids, 3, 659,
  \dodoi{10.1063/1.1706104}

\bibitem[{{Sprouse} {et~al.}(2020){Sprouse}, {Mumpower}, \&
  {Surman}}]{Sprouse2020}
{Sprouse}, T.~M., {Mumpower}, M.~R., \& {Surman}, R. 2020, in European Physical
  Journal Web of Conferences, Vol. 242, European Physical Journal Web of
  Conferences, 04001, \dodoi{10.1051/epjconf/202024204001}

\bibitem[{{Sprouse} {et~al.}(2021){Sprouse}, {Mumpower}, \&
  {Surman}}]{Sprouse2021}
{Sprouse}, T.~M., {Mumpower}, M.~R., \& {Surman}, R. 2021, \prc, 104, 015803,
  \dodoi{10.1103/PhysRevC.104.015803}

\bibitem[{{Suzuki} \& {Maeda}(2022)}]{Suz22}
{Suzuki}, A., \& {Maeda}, K. 2022, \apj, 925, 148,
  \dodoi{10.3847/1538-4357/ac3d8d}

\bibitem[{Suzuki {et~al.}(2010)Suzuki, Sato, \& Lee}]{Suzuki2010}
Suzuki, N., Sato, T., \& Lee, T.-S.~H. 2010, Phys. Rev. C, 82, 045206,
  \dodoi{10.1103/PhysRevC.82.045206}

\bibitem[{Taub(1948)}]{Taub1948}
Taub, A.~H. 1948, Phys. Rev., 74, 328, \dodoi{10.1103/PhysRev.74.328}

\bibitem[{{Usov}(1992)}]{Usov92}
{Usov}, V.~V. 1992, \nat, 357, 472, \dodoi{10.1038/357472a0}

\bibitem[{Valera {et~al.}(2022)Valera, Bustamante, \& Glaser}]{Valera2022}
Valera, V.~B., Bustamante, M., \& Glaser, C. 2022, Journal of High Energy
  Physics, 2022, \dodoi{10.1007/jhep06(2022)105}

\bibitem[{{Vassh} {et~al.}(2024){Vassh}, {Wang}, {Larivi{\`e}re}, {Sprouse},
  {Mumpower}, {Surman}, {Liu}, {McLaughlin}, {Denissenkov}, \&
  {Herwig}}]{Vassh2024}
{Vassh}, N., {Wang}, X., {Larivi{\`e}re}, M., {et~al.} 2024, \prl, 132, 052701,
  \dodoi{10.1103/PhysRevLett.132.052701}

\bibitem[{{Vurm} {et~al.}(2011){Vurm}, {Beloborodov}, \& {Poutanen}}]{Vurm11}
{Vurm}, I., {Beloborodov}, A.~M., \& {Poutanen}, J. 2011, \apj, 738, 77,
  \dodoi{10.1088/0004-637X/738/1/77}

\bibitem[{Wambach(2003)}]{Wambach2003}
Wambach, J. 2003, Progress in Particle and Nuclear Physics, 50, 615,
  \dodoi{https://doi.org/10.1016/S0146-6410(03)00057-7}

\bibitem[{Wanajo(2007)}]{Wanajo2007}
Wanajo, S. 2007, The Astrophysical Journal, 666, L77–L80,
  \dodoi{10.1086/521724}

\bibitem[{{Wang} {et~al.}(2020){Wang}, {N3AS Collaboration}, {Vassh}, {FIRE
  Collaboration}, {Sprouse}, {Mumpower}, {Vogt}, {Randrup}, \&
  {Surman}}]{Wang2020}
{Wang}, X., {N3AS Collaboration}, {Vassh}, N., {et~al.} 2020, \apjl, 903, L3,
  \dodoi{10.3847/2041-8213/abbe18}

\bibitem[{Waxman \& Bahcall(1997)}]{Waxman1997}
Waxman, E., \& Bahcall, J. 1997, Phys. Rev. Lett., 78, 2292,
  \dodoi{10.1103/PhysRevLett.78.2292}

\bibitem[{Waxman \& Bahcall(1998)}]{Waxman1998}
---. 1998, Physical Review D, 59, \dodoi{10.1103/physrevd.59.023002}

\bibitem[{{Wiescher} {et~al.}(2020){Wiescher}, {deBoer}, {G{\"o}rres}, {Gula},
  \& {Liu}}]{Wiescher2020}
{Wiescher}, M., {deBoer}, R.~J., {G{\"o}rres}, J., {Gula}, A., \& {Liu}, Q.
  2020, Acta Physica Polonica B, 51, 631, \dodoi{10.5506/APhysPolB.51.631}

\bibitem[{Wietfeldt \& Greene(2011)}]{Wietfeldt2011}
Wietfeldt, F.~E., \& Greene, G.~L. 2011, Rev. Mod. Phys., 83, 1173,
  \dodoi{10.1103/RevModPhys.83.1173}

\bibitem[{{Woosley}(1993)}]{Woos93}
{Woosley}, S.~E. 1993, \apj, 405, 273, \dodoi{10.1086/172359}

\bibitem[{{Woosley} \& {Heger}(2006)}]{WH06}
{Woosley}, S.~E., \& {Heger}, A. 2006, \apj, 637, 914, \dodoi{10.1086/498500}

\bibitem[{Woosley {et~al.}(2002)Woosley, Heger, \& Weaver}]{Woosley2002}
Woosley, S.~E., Heger, A., \& Weaver, T.~A. 2002, Rev. Mod. Phys., 74, 1015,
  \dodoi{10.1103/RevModPhys.74.1015}

\bibitem[{{Yang} {et~al.}(2018){Yang}, {Kafexhiu}, \& {Aharonian}}]{Yang2018}
{Yang}, R.-z., {Kafexhiu}, E., \& {Aharonian}, F. 2018, \aap, 615, A108,
  \dodoi{10.1051/0004-6361/201730908}

\bibitem[{{Yang} {et~al.}(2024){Yang}, {Troja}, {O'Connor}, {Fryer}, {Im},
  {Durbak}, {Paek}, {Ricci}, {Bom}, {Gillanders}, {Castro-Tirado}, {Peng},
  {Dichiara}, {Ryan}, {van Eerten}, {Dai}, {Chang}, {Choi}, {De}, {Hu},
  {Kilpatrick}, {Kutyrev}, {Jeong}, {Lee}, {Makler}, {Navarete}, \&
  {P{\'e}rez-Garc{\'\i}a}}]{Yang2024}
{Yang}, Y.-H., {Troja}, E., {O'Connor}, B., {et~al.} 2024, \nat, 626, 742,
  \dodoi{10.1038/s41586-023-06979-5}

\bibitem[{{Zhang} \& {M{\'e}sz{\'a}ros}(2004)}]{ZM04}
{Zhang}, B., \& {M{\'e}sz{\'a}ros}, P. 2004, International Journal of Modern
  Physics A, 19, 2385, \dodoi{10.1142/S0217751X0401746X}

\bibitem[{{Zhang} \& {Yan}(2011)}]{ZY11}
{Zhang}, B., \& {Yan}, H. 2011, \apj, 726, 90,
  \dodoi{10.1088/0004-637X/726/2/90}

\bibitem[{{Zhu} {et~al.}(2018){Zhu}, {Wollaeger}, {Vassh}, {Surman}, {Sprouse},
  {Mumpower}, {M{\"o}ller}, {McLaughlin}, {Korobkin}, {Kawano}, {Jaffke},
  {Holmbeck}, {Fryer}, {Even}, {Couture}, \& {Barnes}}]{Zhu2018}
{Zhu}, Y., {Wollaeger}, R.~T., {Vassh}, N., {et~al.} 2018, \apjl, 863, L23,
  \dodoi{10.3847/2041-8213/aad5de}

\bibitem[{{Zhu} {et~al.}(2021){Zhu}, {Lund}, {Barnes}, {Sprouse}, {Vassh},
  {McLaughlin}, {Mumpower}, \& {Surman}}]{Zhu2021}
{Zhu}, Y.~L., {Lund}, K.~A., {Barnes}, J., {et~al.} 2021, \apj, 906, 94,
  \dodoi{10.3847/1538-4357/abc69e}

\end{thebibliography}

\end{document}